\documentclass[aps,pra,amsmath,amssymb,twocolumn,superscriptaddress,10pt,notitlepage,nofootinbib,longbibliography]{revtex4-1}

\graphicspath{ {./images/} }
\begin{document}

\title{Entanglement-enhanced time-continuous quantum control in optomechanics}
\author{Sebastian G.\ Hofer}
\email{sebastian.hofer@univie.ac.at}
\affiliation{Vienna Center for Quantum Science and Technology (VCQ), Faculty of Physics, University of Vienna, Boltzmanngasse 5, 1090 Vienna, Austria}
\affiliation{Institute for Theoretical Physics, Institute for Gravitational Physics (Albert Einstein Institute), Leibniz University Hannover, Callinstra\ss{}e 38, 30167 Hannover, Germany}
\author{Klemens Hammerer}
\affiliation{Institute for Theoretical Physics, Institute for Gravitational Physics (Albert Einstein Institute), Leibniz University Hannover, Callinstra\ss{}e 38, 30167 Hannover, Germany}

\begin{abstract}
  The cavity-optomechanical radiation pressure interaction provides the means to create entanglement between a mechanical oscillator and an electromagnetic field interacting with it. Here we show how we can utilize this entanglement within the framework of time-continuous quantum control, in order to engineer the quantum state of the mechanical system. Specifically, we analyze how to prepare a low-entropy mechanical state by (measurement-based) feedback cooling operated in the blue detuned regime, the creation of bipartite mechanical entanglement via time-continuous entanglement swapping, and preparation of a squeezed mechanical state by time-continuous teleportation. The protocols presented here are feasible in optomechanical systems exhibiting a cooperativity larger than 1.
\end{abstract}
\maketitle

\section{Introduction}
\label{sec-1}

Quantum control plays a crucial role in modern quantum experiments across different fields. In optomechanics alone its applications range from feedback cooling of the mechanical motion \cite{cohadon_cooling_1999}, mechanical squeezing \cite{clerk_back-action_2008,woolley_nanomechanical_2008} and two-mode squeezing \cite{woolley_two-mode_2013} to back-action elimination \cite{wiseman_using_1995,courty_quantum_2003} with possible applictions in gravitational wave detection. Importantly for quantum information processing and communication, it can also be used to robustly generate entanglement between remote quantum systems, as has been demonstrated recently for spin qubits \cite{dolde_high-fidelity_2014}. At the same time entanglement itself can be an essential component to facilitate control of quantum systems, \eg{}, as a resource for teleportation \cite{bennett_teleporting_1993} when employed as a means for remote state preparation. In optomechanics, pulsed entanglement between a mechanical oscillator and the electromagnetic field \cite{hofer_quantum_2011} has recently been demonstrated in an electromechanical setup \cite{palomaki_entangling_2013}; state preparation (and verification) of an arbitrary mechanical quantum state (\eg{}, a Fock state) is yet to be accomplished (see, however, \cite{oconnell_quantum_2010}). Typical quantum control protocols are operated in a time-continuous fashion and often rely on continuous measurements which are capable of tracking the quantum state of the controlled system. The resulting measurement record---and the so-called conditional quantum state inferred from it---is then used as a basis for the applied feedback \cite{wiseman_quantum_2009}. Thus, the control protocol's success critically depends on the precision of the employed measurement. The most essential prerequisite for quantum limited feedback control turns out to be the regime of strong (linearized, thermal) cooperativity. This regime has been witnessed in several experiments in the past few years \cite{murch_observation_2008,teufel_sideband_2011,chan_laser_2011,brooks_non-classical_2012,safavi-naeini_squeezed_2013,purdy_observation_2013}.
Recently, monitoring a mechanical oscillator with a measurement strength matching its thermal decoherence rate (equivalent to a cooperativity above 1) and measurement-based feedback cooling to an occupation number of several phonons has been demonstrated in \cite{wilson_measurement_2014}.

\begin{figure}[tb]
  \centerline{\includegraphics[width=\columnwidth]{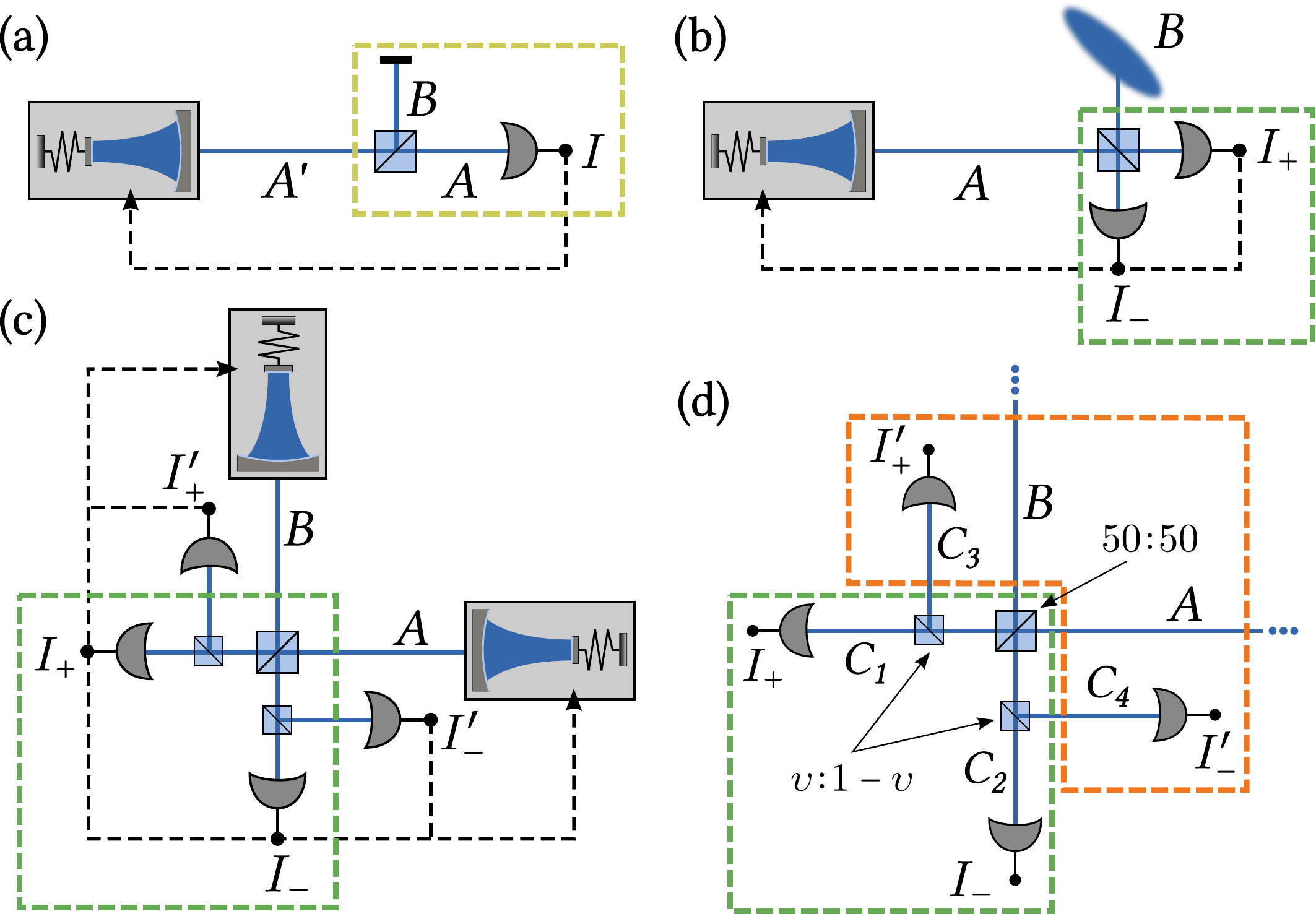}}
  \caption[]{Different optomechanical setups considered in this article: (a) single homodyne detection for feedback cooling (non-unit efficiency is modeled by a beam-splitter before a perfect detector, marked by the yellow box; see \sref{SEC:homodyne-generic}), (b) time-continuous teleportation, (c) time-continuous entanglement swapping, (d) detail of the entanglement swapping setup. The green (dark gray) dashed boxes mark the time-continuous Bell measurement setups, the orange (light gray) box in (d) marks the auxiliar stabilizing measurements.}
  \label{fig:setup}
\end{figure}

In this article we explore protocols which exploit optomechanical entanglement as a resource for measurement-based time-continuous control of cavity-optomechanical systems (see \fref{fig:setup} for the different setups being considered). The presented protocols rely on the fact that for a laser drive appropriately (blue) detuned from the cavity resonance, the radiation-pressure interaction generates entanglement between the mechanical oscillator and the cavity output light. We will show that due to this fact it is in principle possible to feedback-cool the mechanical motion to its ground state, even though the optomechanical interaction effects heating of the mechanical mode in this particular regime. This stands in contrast to common feedback-cooling schemes which operate on cavity resonance \cite{cohadon_cooling_1999,wilson_measurement_2014}. The scheme presented here is based on standard homodyne detection to monitor a single quadrature of the output light, and feedback by (phase and amplitude) modulation of the driving laser. The feedback signal is calculated by applying linear--quadratic--Gaussian (LQG) control from classical control theory, which attains the optimal cooling performance for the chosen configuration. Beyond cooling, entanglement-based protocols for quantum control can be used to achieve more sophisticated quantum state engineering. We study in detail two optomechanical implementations of time-continuous Bell measurements \cite{hofer_time-continuous_2013}: Time-continuous teleportation allows for preparation of a mechanical oscillator in a general Gaussian (squeezed) state, while time-continuous entanglement-swapping can be used to prepare two remote mechanical systems in an (Einstein--Podolsky--Rosen) entangled state. Both schemes generate dissipative dynamics which drive the mechanical system(s) into the desired stationary state. They are shown to work if the effective measurement strength is on the same order as the mechanical decoherence rate (\ie{}, for a optomechanical cooperativity of around 1), which is the same condition that holds for ground-state cooling \cite{teufel_sideband_2011,chan_laser_2011}, observation of back-action noise \cite{murch_observation_2008,purdy_observation_2013}, and ponderomotive squeezing \cite{brooks_non-classical_2012,safavi-naeini_squeezed_2013}, all of which have been achieved experimentally. Although we here consider optomechanical systems only, the presented methods are very versatile, applicable to different (continuous and discrete) physical systems, and can be extended to describe more complex topologies, such as multiple interferometric measurements and quantum networks \cite{hofer_time-continuous_2013}.

The results on time-continuous teleportation and entanglement swapping have been published in parts in \cite{hofer_time-continuous_2013}. Here we provide an extended treatment focusing on an optomechanical implementation and present a detailed derivation of the resulting equations.

The manuscript is organized as follows: In \sref{SEC:central} we summarize and illustrate the central results concerning cooling, mechanical squeezing, and bipartite mechanical entanglement generation. We start this section by discussing the phase diagram of the optomechanical steady state, emphasizing its unique features for blue detuned laser drive. \sref{SEC:meq} presents in detail all technical aspects in the derivation of our results. Some background information about quantum stochastic calculus and LQG control is presented in the appendices \ref{APP:qsc} and \ref{APP:LQG}.

\section{Central results}
\label{sec-2}
\label{SEC:central}

\subsection{The cavity-optomechanical system}
\label{sec-2-1}
\label{SEC:omsys}

In this article we consider a cavity optomechanical system with a single mechanical mode oscillating at a resonance frequency \(\om\). The cavity has a resonance frequency \(\omega_c\) and a [full width at half maximum (FWHM)] decay rate \(\kappa\), and is driven by continuous-wave laser light at a frequency \(\omega_l\).
In a linearized description and in a frame rotating with $\omega_l$, the system is described by the effective Hamiltonian \cite{aspelmeyer_cavity_2014}
\begin{equation}
  \label{eq:1}
  H_{\mathrm{sys}}=\om\cm^{\dagger}\cm - \Dc\cc^{\dagger}\cc + g(\cm+\cm^{\dagger})(\cc+\cc^{\dagger}),
\end{equation}
where \(\cm\) and \(\cc\) are bosonic annihilation operators of the mechanical and the optical mode respectively. \(\Dc=\omega_0-\omega_{\mathrm{c}}\) is the detuning of the driving laser with respect to the cavity, and \(g\) is the optomechanical coupling strength. In writing this Hamiltonian we implicitly assumed that the cavity is driven strongly, such that the radiation-pressure interaction can be linearized around a large classical intracavity amplitude. The coupling strength is then given by \(g=g_0[2\kappa P/\hbar\omega_0(\kappa^2+\Delta^2)]^{1/2}\) with the single-photon coupling \(g_0\) and the input laser power \(P\). To work with the linearized description we assume the existence of a {unique} classical steady state with a large intracavity photon number, thus neglecting effects of bistability \cite{ghobadi_quantum_2011,aspelmeyer_cavity_2014}. Additionally we assume $g_0\ll \kappa, \om$, which is needed to safely neglect nonlinear radiation pressure effects \cite{aspelmeyer_cavity_2014}.

The linearized radiation-pressure Hamiltonian \(H_{\mathrm{om}}=g(\cm+\cm^{\dagger})(\cc+\cc^{\dagger})\) can be decomposed into two terms: a beam-splitter like interaction \(g(\cm\cc^{\dagger}+\cm^{\dagger}\cc)\) which is resonant for \(\Dc=-\om\) (red detuned laser drive) and effects cooling of the mechanical motion, and a two-mode squeezing term \(g(\cm\cc+\cm^{\dagger}\cc^{\dagger})\) which is dominant for \(\Dc=\om\) (blue driving) and is responsible for creation of optomechanical correlations and entanglement. For a resonant drive (\(\Dc=0\)) we retain the full Hamiltonian \(\propto g x_{\m}x_{\lm}\), which is commonly associated with quantum non-demolition (QND) measurements of harmonic oscillators \cite{thorne_quantum_1978,braginsky_quantum_1980} and is the interaction typically used for measurement-based feedback control of these systems \cite{mancini_optomechanical_1998,doherty_feedback_1999,hamerly_coherent_2013}.
The optical and mechanical quadrature operators we define  as \(x_i=(c_i+c_i^{\dagger})/\sqrt{2}\) and \(p_i=-\ii(c_i-c_i^{\dagger})/\sqrt{2}\) (\(i\in \left\{ \m,\lm \right\}\)) which leads to canonical commutation relations \([x_i,p_j]=\ii\delta_{ij}\). In the following it will be convenient to collect them into the vector operator \(\vc{X}=(x_{\m},p_{\m},x_{\lm},p_{\lm})^{\trans}\).

\subsection{The optomechanical phase diagram}
\label{sec-2-2}

Optomechanical sideband cooling and entanglement creation in steady state have been analyzed in detail in the literature \cite{wilson-rae_theory_2007,marquardt_quantum_2007,genes_ground-state_2008,genes_robust_2008}. Both phenomena are captured by the standard optomechanical master equation (MEQ) for the quantum state \(\rho\), given by \cite{wilson-rae_theory_2007}
\begin{equation}
  \label{eq:2}
  \begin{split}
    \dot{\rho}(t) &=\mathcal{L}\rho(t)=-\ii[H,\rho(t)]+\kappa\mathcal{D}[\cc]\rho(t)\\
    &\qquad+\gamma(\bar{n}+1)\mathcal{D}[\cm]\rho(t)+\gamma\bar{n}\mathcal{D}[\cm^{\dagger}]\rho(t)
  \end{split}
\end{equation}
where \(\mathcal{L}\) is the so-called Liouville operator. Here \(\gamma\) denotes the (FWHM) width of the mechanical resonance, and \(\bar{n}\) the mechanical bath's mean phonon number. Optical and mechanical decoherence is described by the Lindblad operators \(\mathcal{D}[c]\rho=c\rho c^{\dagger}-\tfrac{1}{2}\rho c^{\dagger}c-\tfrac{1}{2}c^{\dagger}c\rho\). As our system is Gaussian, its state is fully characterized by the first and second moments of \(\vc{X}\), \ie{}, the mean values \(\mean{\vc{X}}(t)=\tr{\vc{X}\rho(t)}\) and the symmetric covariance matrix
\begin{equation}
  \label{eq:3}
  \mat{\Sigma}(t)=\Re \bigr( \mean{\vc{X} \vc{X}^{\trans}}(t) \bigl)-\mean{\vc{X}}(t)\mean{\vc{X}^{\trans}}(t).
\end{equation}
(Throughout this paper we will often omit the explicit time argument for the sake of brevity if no ambiguity exists.)
The linear equations of motion of \(\mean{\vc{X}}\) and \(\mat{\Sigma}\) are given by (\aref{APP:LQG})
\begin{subequations}
  \label{eq:4}
  \begin{align}
    \label{eq:5}
    \frac{\dd{}}{\dt}\mean{\vc{X}}(t)&=\mat{F}\mean{\vc{X}}(t),\\
    \label{eq:6}
    \frac{\dd{}}{\dt}\mat{\Sigma}(t)&=\mat{F}\mat{\Sigma}(t)+\mat{\Sigma}(t) \mat{F}^{\trans}+\mat{N}.
  \end{align}
\end{subequations}
The \(4\times 4\) matrices \(\mat{F}\) and \(\mat{N}\) describe the system's dynamics and noise properties respectively, and are algebraically connected to the Liouvillian in the MEQ \eqref{eq:2}.

Provided the system is stable, it will in the long-term assume a steady state, \(\lim_{t\rightarrow \infty} \rho(t)=\sss{\rho}\), where \(\sss{\rho}\) is determined by the condition \(\mathcal{L}\sss{\rho}=0\).
The stability of a linear system can be assessed by applying the Routh--Hurwitz (RH) stability criterion \cite{i.s._gradshteyn_table_2007} which is fulfilled iff all eigenvalues of \(\mat{F}\) have negative real parts. If a stable steady state exists the means \(\sss{\mean{\vc{X}}}{:=}\tr{\vc{X}\sss{\rho}}=0\) will vanish, while the steady-state covariance matrix \(\sss{\mat{\Sigma}}\) is given by the solution to the so-called Lyapunov equation, which is obtained from \eqref{eq:6} by setting its left-hand side to zero, \ie{}, \(\mat{F}\sss{\mat{\Sigma}}+\sss{\mat{\Sigma}}\mat{F}^{\trans}+\mat{N}=0\). This equation can readily be solved to obtain steady-state properties such as the mean mechanical occupation number \(\sss{n_{\m}}=\sss{\mean{\cm^{\dagger}\cm}}\) or logarithmic negativity \(E_{\mathcal{N}}^{\mathrm{ss}}\) \cite{vidal_computable_2002,plenio_logarithmic_2005}. In this section we will mainly be concerned with these steady-state properties of optomechanical systems.

\begin{figure}[hb]
  \begin{minipage}{\columnwidth}
    \rlap{(a)}
    \centerline{\includegraphics[width=\columnwidth]{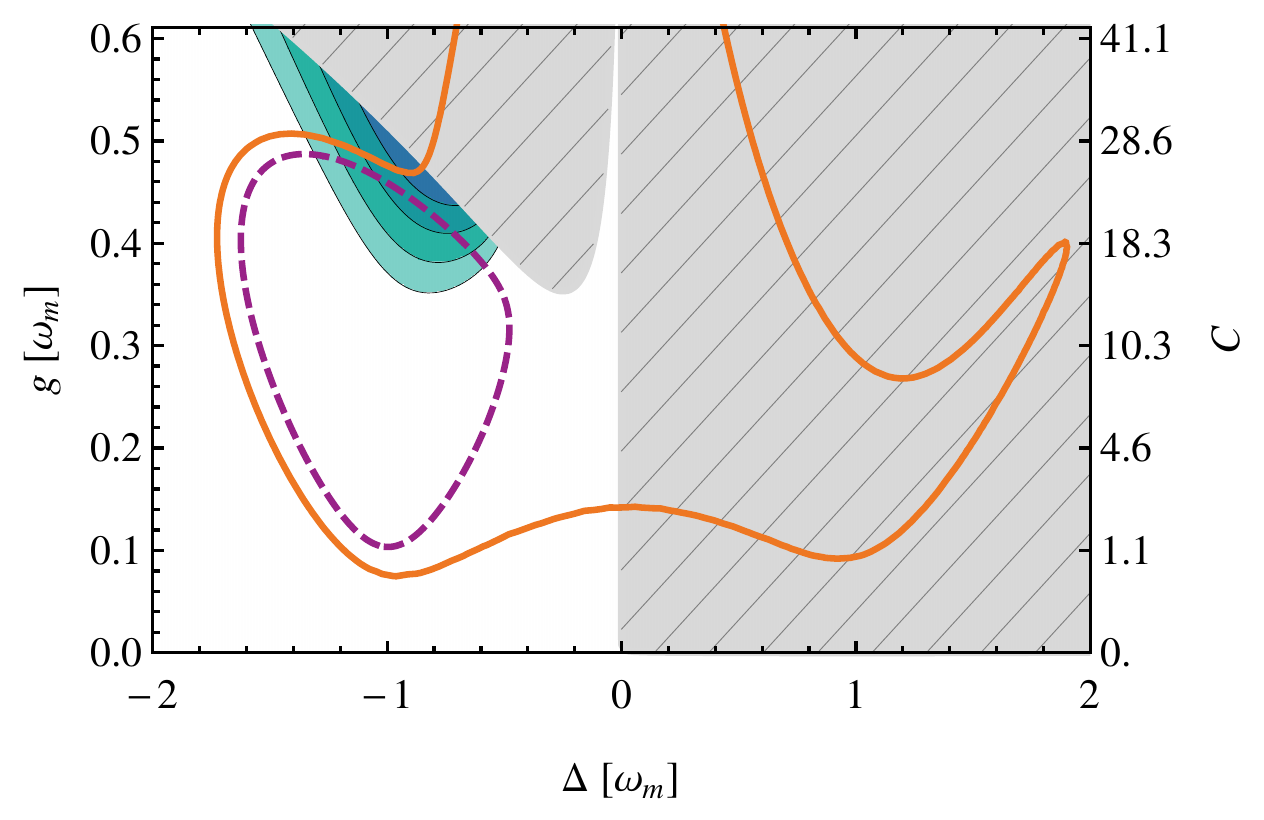}}
  \end{minipage}
  \begin{minipage}{\columnwidth}\rlap{(b)}{\centerline{\includegraphics[width=\columnwidth]{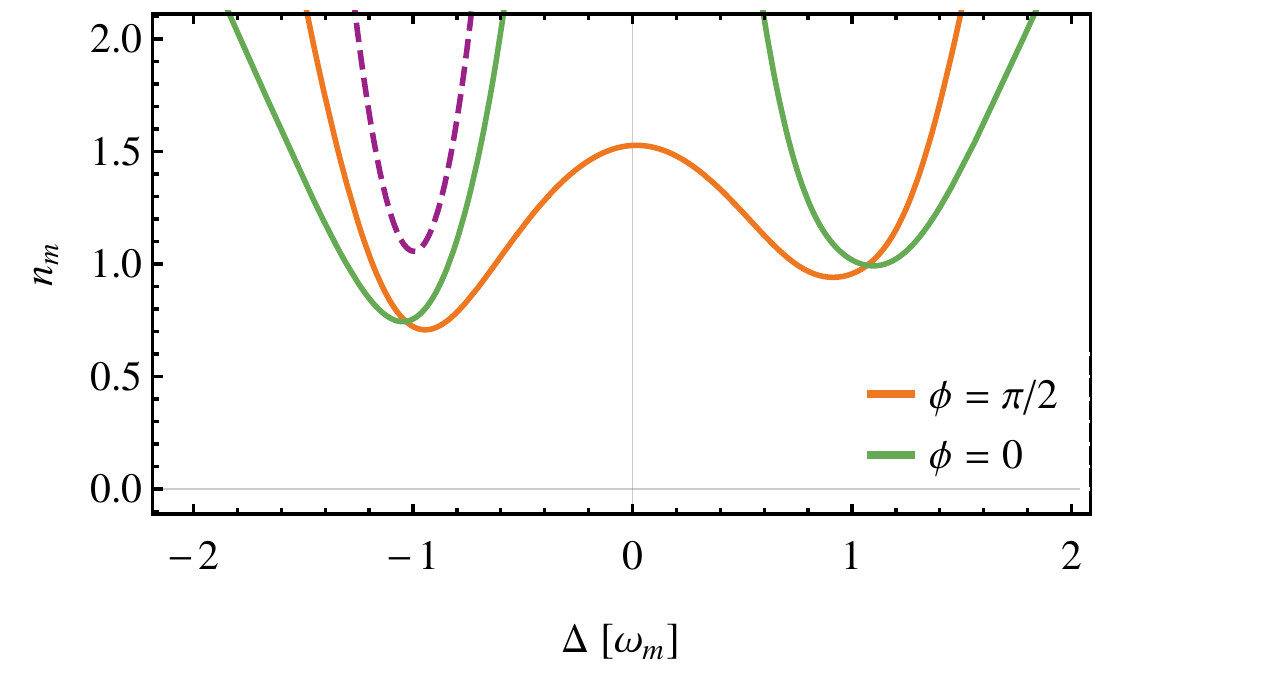}}}\end{minipage}
  \caption[]{(Color online) Upper plot: Steady-state phase diagram of an optomechanical system for $\kappa=\om/2$, $Q=5\cdot 10^6$, $\bar{n}=3.5\cdot 10^5$. The gray hatched area depicts unstable regions where no steady-state exists. The dashed purple (dark gray) line shows regions of ground state cooling of the mechanical oscillator, here $\sss{n_{\m}} < 1$. The turquoise/blue (gray unhatched) area shows optomechanical entanglement (logarithmic negativity $E_{\mathcal{N}}^{\mathrm{ss}}$), with largest values close to the instability region. The orange (light gray) line encloses the regions where the conditional mean mechanical occupation number $\sss{\cond{\mean{c^{\dagger}_{\m}c_{\m}}}}<1$ for a measurement of the optical phase quadrature, \ie{}, $\phi=\pi/2$. The right axis shows the corresponding optomechanical cooperativity, given by $C=4g^2/(\bar{n}+1)\gamma\kappa$.
    Lower Plot: Cut through the phase diagram at $g=\om/10$, depicting the conditional phonon number $\sss{\cond{\mean{c^{\dagger}_{\m}c_{\m}}}}$ for LO phases $\phi=\pi/2$ (orange [light gray] line) and $\phi=0$ (green [dark gray] line). The dashed purple line again shows the mean occupation number for the unconditional state for sideband cooling.
  }
  \label{fig:phasediag}
\end{figure}

The characteristic features of an optomechanical system's steady state can nicely be illustrated by plotting a phase diagram with respect to the laser detuning \(\Dc\) and the optomechanical coupling \(g\), as depicted in Fig.~\ref{fig:phasediag} for an optomechanical system in the resolved sideband regime (\(\kappa<\om\)) for a high-Q (\(Q=\om/\gamma\)) mechanical oscillator. The grey background depicts the regions of instability, given by the corresponding Routh--Hurwitz criterion, where no steady state  exists. The first thing to note is that the system is unstable in nearly all the right half-plane, \ie{}, for blue detuned laser drive, while for red detuning the system becomes unstable only for appreciably high optomechanical coupling. Centered around the first mechanical sideband at \(\Dc=-\om\) where the beam-splitter part of the optomechanical interaction is resonant, lies the region where \(\sss{n_{\m}}<1\) (dashed purple line) and thus ground-state cooling is possible. Right at the border of stability, for a similar detuning, we find regions of large steady-state entanglement between the intracavity field and the mechanical resonator (colored in turquoise/blue) \cite{genes_robust_2008}.  On the opposite side of the phase diagram, around the blue mechanical sideband at \(\Dc=\om\), we also expect to observe optomechanical entanglement due to the effect of the optomechanical two-mode squeezing dynamics. However, there the formation of a steady state is inhibited by the optomechanical instability which is due to parametric amplification of the amplitude of both the mechanical and the optical mode \cite{aspelmeyer_cavity_2014}. The connection of laser cooling, entanglement generation, and the instability region has been analyzed in detail in \cite{genes_robust_2008}.
Although no steady state exists for a blue-detuned laser drive, various alternative approaches permit to work with the resonantly enhanced two-mode squeezing dynamics of the optomechanical interaction. Pulsed optomechanical entanglement creation, for example,---which does not require to be operated in a stable regime---has been analyzed in detail in \cite{hofer_quantum_2011}, and has been experimentally demonstrated (for an electromechanical system) in \cite{palomaki_entangling_2013}. Working with a continuous-wave blue-detuned laser drive on the other hand, is still possible if we employ stabilizing feedback which inhibits the exponential growth of the optomechanical system's quadratures. One possible type of feedback is measurement-based feedback using homodyne detection, which we will consider in the following.

By adding a homodyne detector to our setup [measuring a single light quadrature defined by the local oscillator (LO) angle \(\phi\)], we can condition the optomechanical system's state on the resulting photo-current \(I(t)\), which leads to a stochastic master equation (SME) in the \Ito{} sense (see \cite{gardiner_quantum_2004,wiseman_quantum_2009} and \sref{SEC:homodyne-generic})
\begin{equation}
  \label{eq:7}
  \dd\cond{\rho}(t) = \mathcal{L}\cond{\rho}(t)\dt + \sqrt{\eta\kappa}\,\mathcal{H}[\ee^{\ii \phi}\cc]\cond{\rho}(t)\dW(t).
\end{equation}
\(\cond{\rho}\) is the so-called \emph{conditional quantum state}, which describes our knowledge of the system given a specific measurement record \(I(t)\). The effect of conditioning is described by the operator \(\mathcal{H}[s]\cond{\rho}=(s-\tr{s\cond{\rho}})\cond{\rho}+\cond{\rho}(s-\tr{s\cond{\rho}})^{\dagger}\). \(\mathcal{H}\) is thus nonlinear in \(\cond{\rho}\), as is expected for a measurement term. \(I(t)\) can be expressed as
\begin{equation}
  \label{eq:8}
  I(t)\dt=\sqrt{\eta\kappa}\cond{\mean{\cc\ee^{-\ii\phi}+\cc^{\dagger}\ee^{\ii\phi}}}(t)\dt+\dW(t),
\end{equation}
where \(\dd{W}\) is a Wiener increment with \(\dW(t)^2=\dt\), and \(0<\eta<1\) is the efficiency of the detection. Here and in the following we denote by \(\cond{\mean{A}}(t)=\tr{A\cond{\rho}(t)}\) the expectation value with respect to the conditional state. In contrast to the conditional state \(\cond{\rho}\) which solves a SME, we will call the solution of a standard MEQ [such as \eqref{eq:2}] the \emph{unconditional state}, which we denote by \(\rho\).

\begin{figure}[htb]
  \centerline{\includegraphics[width=.75\columnwidth]{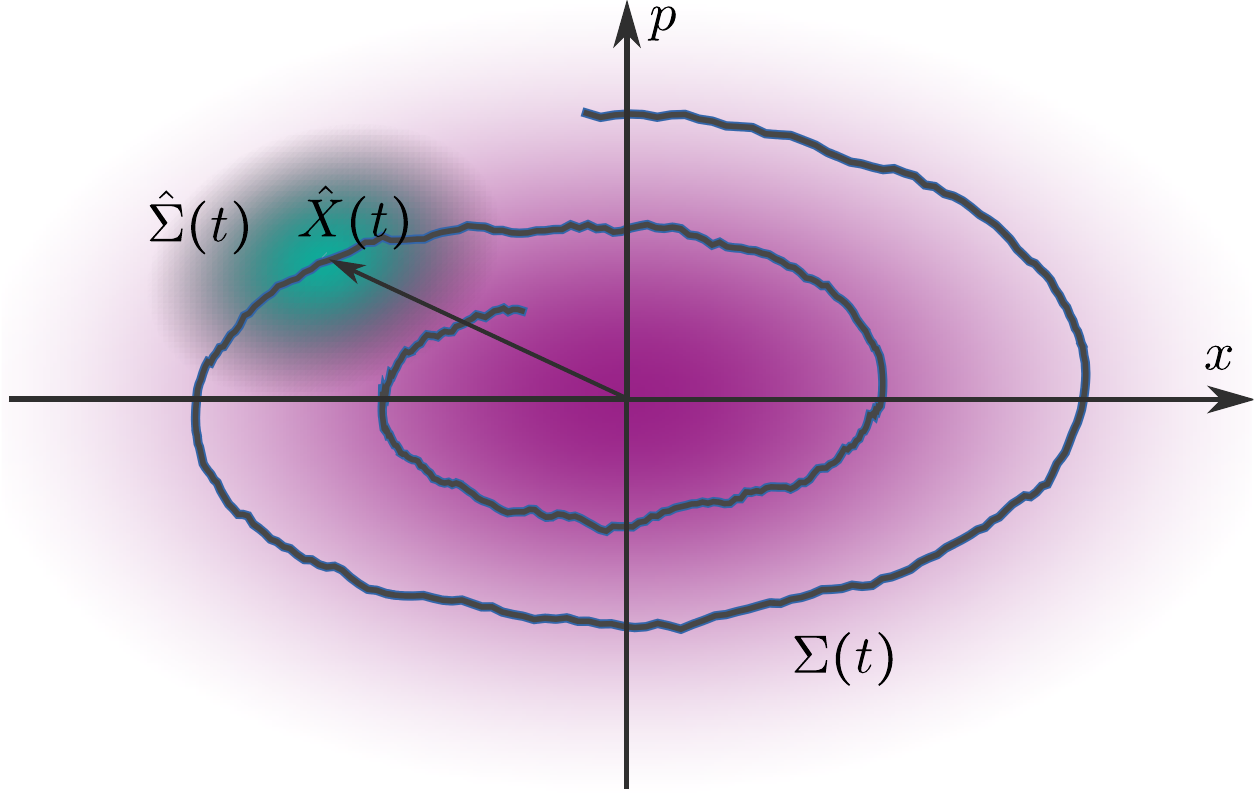}}
  \caption[]{(Color online) Schematic comparison of the master equation \eqref{eq:2} and the stochastic master equation \eqref{eq:7} for a single mode Gaussian system in phase space. The conditional state with a covariance $\mat{\est{\Sigma}}(t)$, depicted by the turquoise  (small, off-centered) elipse, moves through phase space on a trajectory given by a realisation of $\vc{\est{X}}(t)$ (blue [dark gray] line). Averaging over many sample paths recovers the broad, unconditional distribution, determined by $\mat{\Sigma}(t)$ (purple [large, centered] ellipse).}
  \label{fig:phasespace}
\end{figure}

Gaussian conditional quantum states are fully described by the mean vector \(\vc{\hat{X}}(t)=\cond{\mean{\vc{X}}}(t)\) and covariance matrix
\begin{equation}
  \label{eq:9}
  \mat{\est{\Sigma}}(t)=\Re \bigr( \cond{\mean{\vc{X} \vc{X}^{\trans}}}(t) \bigl)-\cond{\mean{\vc{X}}}(t)\cond{\mean{\vc{X}^{\trans}}}(t),
\end{equation}
defined with respect to \(\cond{\rho}\). Their equations of motion are given by a linear stochastic differential equation and a (deterministic) matrix Riccati equation respectively,
\begin{align}
  \label{eq:10}
  \dd{\vc{\est{X}}}(t)&= \mat{F}\vc{\est{X}}(t)\dt+\mat{K}(t)\bigl[I(t)-\mat{H}\vc{\est{X}}(t)\bigr]\dt,\\
  \label{eq:11}
  \frac{\dd{}}{\dt}{\mat{\est{\Sigma}}(t)}&= \mat{F}\mat{\est{\Sigma}}(t)+\mat{\est{\Sigma}}(t) \mat{F}^{\trans}+\mat{N}\notag\\
                      &\qquad-\big[\mat{\est{\Sigma}}(t) \mat{H}^{\trans}+\mat{M}\big]\big[\mat{\est{\Sigma}}(t) \mat{H}^{\trans}+\mat{M}\big]^{\trans},
\end{align}
where \(\mat{H}\) describes the homodyne measurement and \(\mat{M}\) is related to the system's noise properties (see \aref{APP:LQG}). \(\mat{K}(t)\) is a time-dependent gain factor which depends on \(\mat{\est{\Sigma}}(t)\).
For a one-dimensional system [with a two-dimensional phase space \((x,p)\)] these equations allow us to give a simple graphic interpretation of the SME \eqref{eq:7} in terms of a phase-space description (see \fref{fig:phasespace}): The conditional trajectory \(\vc{\est{X}}\) (blue line) is determined by the measurements \(I(t)\) and therefore follows a random walk in phase space. The covariance matrix \(\mat{\est{\Sigma}}\) (turquoise ellipse) on the other hand evolves deterministically, independent of the measurement results. Averaging over all possible phase-space trajectories recovers the broad Gaussian distribution described by the standard MEQ \eqref{eq:2} [or equivalently, equations \eqref{eq:4}]. For an unstable system (\eg{}, in the blue detuned regime), the blue line will spiral outwards, leading to a growing unconditional covariance. The conditional covariance matrix \(\mat{\est{\Sigma}}\), however, may still possess a (finite) steady state. This is due to the fact that the exponential growth is tracked by the conditional mean, with respect to which the covariance matrix is defined. The steady-state conditional covariance matrix \(\sss{\mat{\est{\Sigma}}}\) can be found in analogy to \(\sss{\mat{\Sigma}}\) by setting the left-hand side of equation \eqref{eq:11} to zero and by solving the resulting algebraic Riccati equation \cite{wiseman_quantum_2009}.

Having obtained \(\sss{\mat{\est{\Sigma}}}\) we can easily evaluate the conditional state's mean mechanical occupation number, depicted in \fref{fig:phasediag}(a) for a measurement of the phase quadrature of the light field, \ie{}, \(\phi=\pi/2\). We find a large region where \(\sss{\cond{\mean{c^{\dagger}_{\m}c_{\m}}}}<0.4\) for all detunings \(-\om\lesssim \Dc \lesssim \om\) (orange line). In the region around the red sideband \(\Dc\approx -\om\) this effect can mainly be attributed to passive sideband cooling of the mirror, which we discussed above. However, we now also find a region of low occupation on the opposite (blue) sideband at \(\Dc\approx \om\). In this region the reduction of the conditional phonon number, which at the same time means an increase of the mechanical state's purity, is due to correlations between the mechanical oscillator and the light field. These correlations allow us to extract information about the mechanics from the homodyne measurement. We will see in the next section that in the sideband-resolved regime \(\kappa<\om\) this effect is strongest for \(\Dc\approx \om\) where the two-mode squeezing, entangling term of the optomechanical Hamiltonian is resonant.

To illustrate how the choice of the LO phase influences the conditional mechanical occupation we plot a cut through \fref{fig:phasediag}(a) at a fixed optomechanical coupling \(g=\om/20\) in \fref{fig:phasediag}(b). If we choose to measure the optical amplitude quadrature we find that on resonance we do not have a reduction of the conditional phonon number. For a detuned laser drive (\(|\Dc|\gtrsim \om\)) however, we again find regions of \(\sss{\cond{\mean{c^{\dagger}_{\m}c_{\m}}}}<1\). This is easily explained by noting that on resonance (\(\Dc=0\)) only the optical phase quadrature couples to the mechanical oscillator, while the amplitude quadrature contains noise only. Measuring the amplitude quadrature therefore does not allow us to make inferences about the mechanical motion. In general there will be an optimal LO angle, depending on all system parameters (especially \(g\), \(\Dc\), \(\kappa\)) at which we obtain maximal information about the mechanical motion. Thus, homodyne detection at this particular angle yields the minimal conditional occupation. Typically---especially in the weak-coupling regime where \(g<\kappa\)---the optimal angle corresponds to the optical quadrature which is \emph{anti}-squeezed by the optomechanical interaction and thus features the best signal-to-noise ratio. We will see in the next section how these features of the optomechanical phase diagram connect to feedback cooling of the mechanical oscillator.

\subsection{Optomechanical Feedback Cooling}
\label{sec-2-3}
\label{SEC:cooling}
Now that we discussed the conditional optomechanical state in detail, the question arises whether it can be realized via feedback, \ie{}, whether we can prepare the \emph{unconditional} state of the system such that it resembles the conditional state.

Consider the setup depicted in \fref{fig:setup}(a), where the results from a homodyne measurement of the cavity output light are fed back to the optomechanical system in a suitable manner, such that the mechanical system is driven to a low entropy steady state. This situation has been analyzed in \cite{mancini_optomechanical_1998,doherty_feedback_1999,genes_ground-state_2008,hamerly_coherent_2013}. However, the regime discussed for feedback cooling is typically restricted to resonant drive and the bad-cavity regime \(\kappa>\om\). In this section we will discuss that feedback cooling can also be effectively operated in the sideband-resolved regime \(\kappa<\om\), and even on the blue sideband \(\Dc=\om\), which is normally affiliated with heating. Here we will show that we can harness the entanglement created by the optomechanical two-mode squeezing interaction for a measurement-based feedback scheme, which enables us to cool the mechanical motion to its ground state.

Feedback onto the mechanical system can either be effected by direct driving through a piezoelectric device \cite{poggio_feedback_2007}, or by modulation of the laser input, as we will assume in the following. This type of optical feedback can be described by adding an additional time-dependent term
\begin{equation}
  \label{eq:12}
  H_{\mathrm{fb}}=-\ii \sqrt{\kappa}[\varepsilon(t)^{\conj}\cc-\varepsilon(t)\cc^{\dagger}]=\sqrt{2\kappa}[u_p(t)x_{\lm}+u_x(t)p_{\lm}]
\end{equation}
to the Hamiltonian, where \(\varepsilon(t)=u_x(t)+\ii u_p(t) \in \mathbb{C}\) is the complex amplitude of the feedback signal, and \(|\varepsilon(t)|^2\) accordingly is the incoming photon flux. To choose an appropriate feedback strategy we employ quantum linear quadratic Gaussian (LQG) control \cite{belavkin_measurement_1999}, which is designed to minimize a quadratic cost function as described in \aref{APP:LQG}. Applied to feedback cooling the basic working principle is the following: From the measurement results of the homodyne detection we calculate the system's conditional state \(\cond{\rho}(t)\), whose evolution is described by \eqref{eq:7}. Based on this state we can then determine the optimal feedback signal \(\varepsilon(t)\) which minimizes the steady-state mechanical occupation number \(\sss{\mean{\cm^{\dagger}\cm}} = \tfrac{1}{2} [\sss{\mean{x_{\m}^2+p_{\m}^2}}-1]\). This of course means that the final occupation number depends on the conditional state (more specifically on the covariance matrix \(\est{\mat{\Sigma}}\)), and thus on the chosen LO angle for the homodyne detection as we discussed above.
A suitable cost function for this problem is given by
\begin{equation}
  \label{eq:13}
  h(x_\m(t),p_\m(t),\varepsilon(t))=h_\m\left[ x_\m(t)^2+p_\m(t)^2\right]+|\varepsilon(t)|^2,
\end{equation}
with \(h_{\m}>0\). Note that \(h\) also includes a contribution by the feedback signal \(\varepsilon\), which precludes feedback strategies with unrealistically high feedback strength. The parameter \(h_{\m}\) therefore effects a trade-off between feedback strength and final occupation number \(\sss{n_{\m}}\): high values of \(h_{\m}\) result in low occupation number possibly requiring large \(|\varepsilon|\) and vice versa. The mean photon flux in the feedback signal can be calculated as described in \aref{APP:LQG}. For the parameters used in this section we find that on average \(|\varepsilon|^2\) is small compared to the overall driving strength in typical experiments. Only in the region of \(\kappa\rightarrow 0\)---where almost no photons enter the cavity---the required \(|\varepsilon|^2\) may increase dramatically. We note that in order for LQG control to work correctly, certain observability and controllability conditions need to be satisfied \cite{wiseman_quantum_2009}, which is indeed the case for our system. Additionally, we assume here that the feedback is instantaneous. In practice this means that any feedback delay $\tau$ should be small on the typical timescales of the system, \ie{}, $\tau \ll 1/\om,\ 1/\kappa$.

\begin{figure}[htb]
  \begin{minipage}{\columnwidth}\rlap{(a)}{\centerline{\includegraphics[width=\columnwidth]{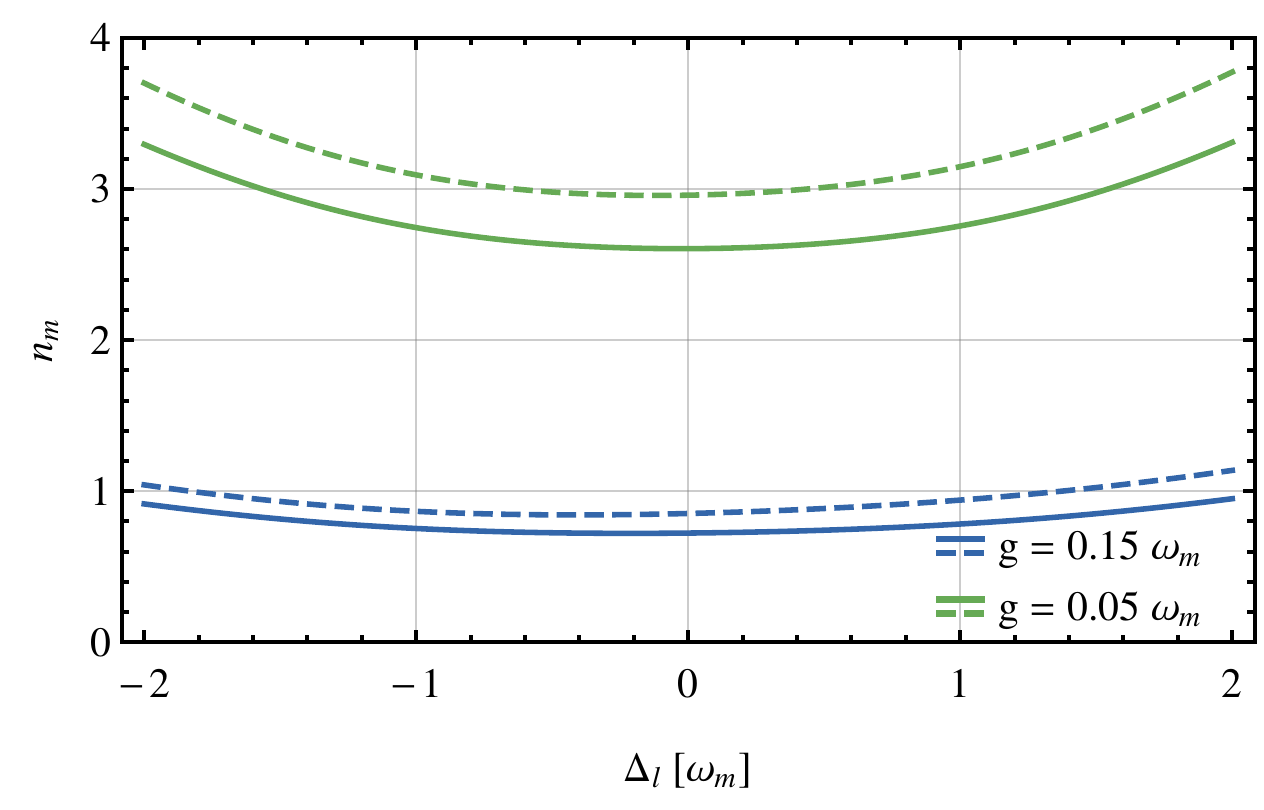}}}\end{minipage}\\
  \begin{minipage}{\columnwidth}\rlap{(b)}{\centerline{\includegraphics[width=\columnwidth]{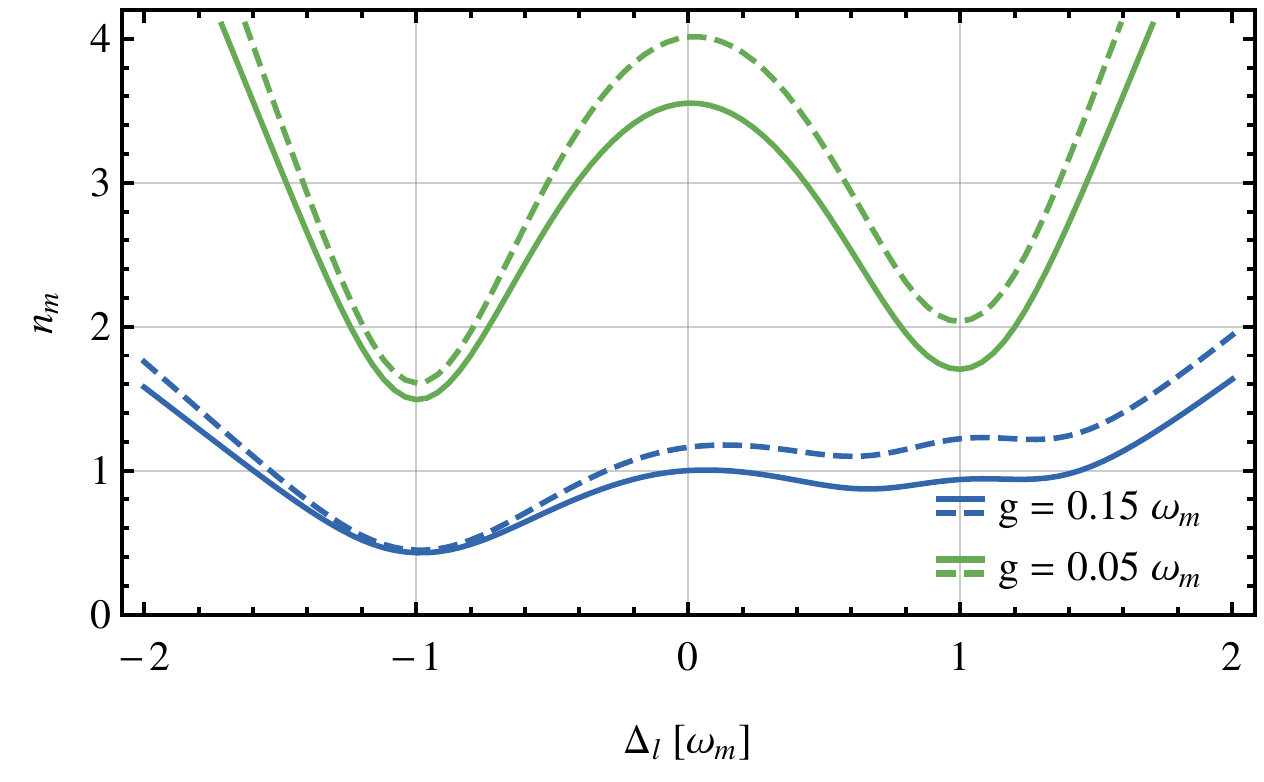}}}\end{minipage}
  \caption[]{(Color online) Steady-state mechanical occupation number $\sss{n_{\m}}$ minimized with respect to the LO angle $\phi$ against detuning of the driving laser $\Dc$ (a) in the bad-cavity regime ($\kappa=4\om$) and (b) in the sideband-resolved regime ($\kappa=\om/2$), for detection efficiencies $\eta=1$ (solid lines) and $\eta=8/10$ (dashed lines). Different colors/gray levels denote different coupling strengths $g$. Other parameters: $Q=5\cdot 10^6$, $\bar{n}=3.5\cdot 10^5$, $h_m=100$}
  \label{fig:occ-det}
\end{figure}
The final mechanical occupation is found by first calculating the steady-state variances \((\Delta x_\m)^2\) and \((\Delta p_\m)^2\) for a closed feedback loop as outlined in \aref{APP:LQG}. \(\sss{n_{\m}}\) is then given by \(\sss{n_{\m}}=\tfrac{1}{2}[(\Delta x_\m)^2+(\Delta p_\m)^2-1]\) (for \(\sss{\mean{x_{\m}}}=\sss{\mean{p_{\m}}}=0\)).

In \fref{fig:occ-det} we plot the steady-state occupation numbers of the feedback-cooled mechanical mode against the laser detuning \(\Dc\), for the bad-cavity regime (upper plot) and the sideband-resolved regime (lower plot), for two different coupling strengths \(g\).  For each detuning the homodyne phase \(\phi\) is chosen such, that the occupation number is minimized.\footnote{This can be achieved in a systematic way by finding the ``optimal unravelling'', see \cite{wiseman_quantum_2009}. Here we simply use a simplex method for optimization.} Note that we keep \(g\) constant while varying \(\Dc\) (or \(\kappa\)). This means that the driving laser power has to be adjusted accordingly.
In the bad-cavity regime \(\kappa>\om\) \com{---the standard regime for feedback cooling---}we find that driving on resonance is favorable for both values of \(g\). In this case the optimal LO phase is \(\phi = \pi/2\), as discussed in the previous section. This is the usual regime for feedback cooling \cite{cohadon_cooling_1999,wilson_measurement_2014}, which is inspired by the idea of quantum non-demolition measurements \cite{braginsky_quantum_1996}, as they are commonly used in gravitational wave detection.

\begin{figure}[htb]
  \begin{minipage}{\columnwidth}\rlap{(a)}{\centerline{\includegraphics[width=\columnwidth]{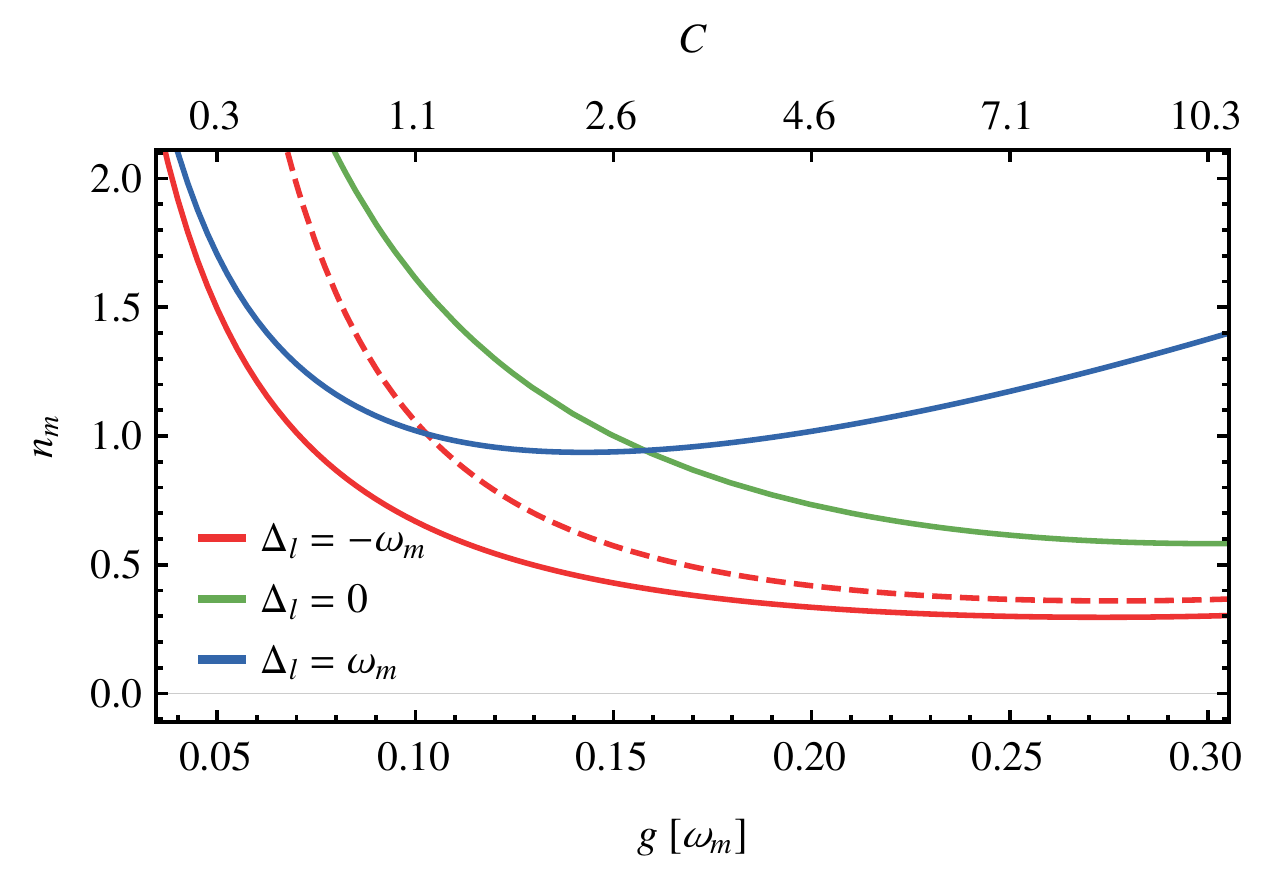}}}\end{minipage}\\
  \begin{minipage}{\columnwidth}\rlap{(b)}{\centerline{\includegraphics[width=\columnwidth]{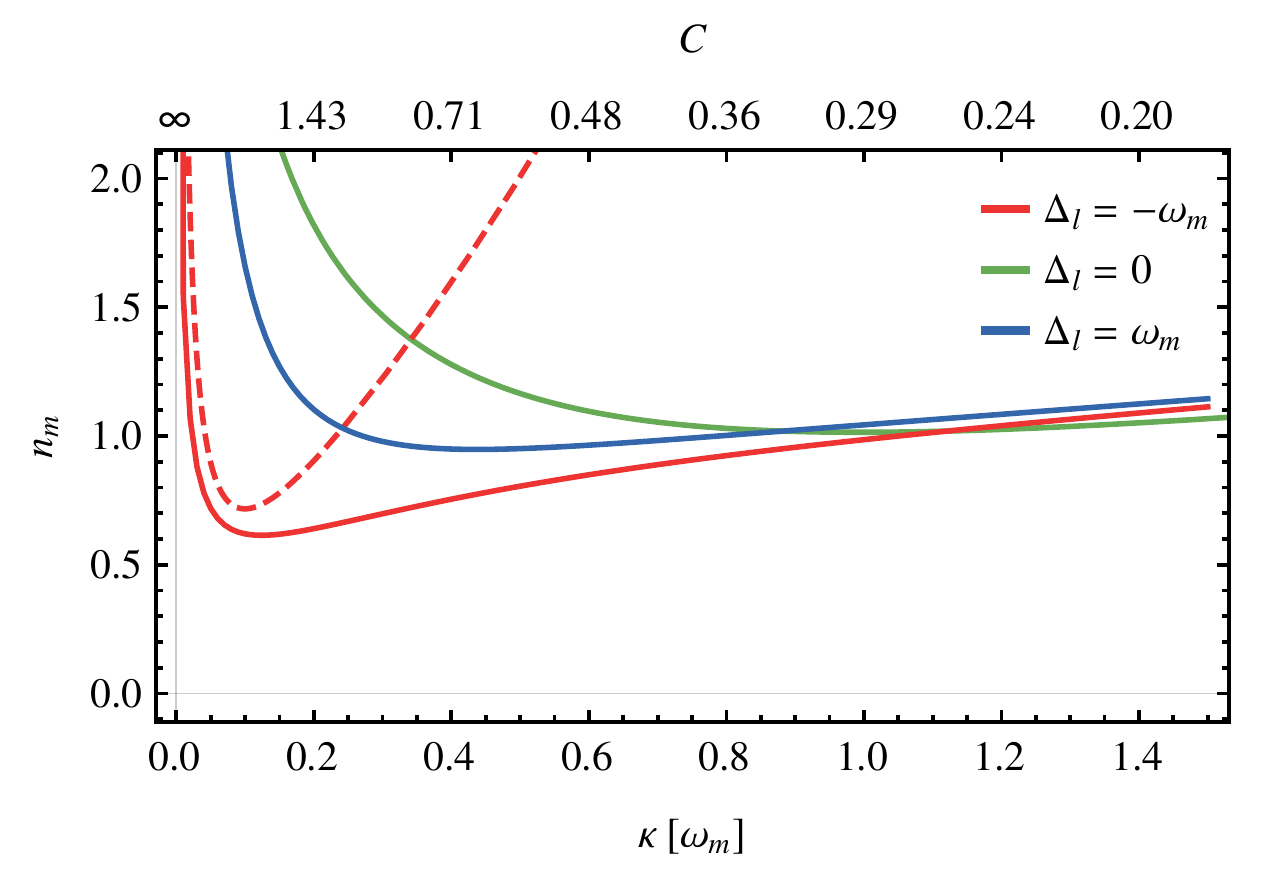}}}\end{minipage}
  \caption[]{(Color online) Steady-state mechanical occupation number $\sss{n_{\m}}$ minimized with respect to $\phi$ for different driving frequencies $\Dc=0,\pm\om$, corresponding to a laser drive on the mechanical sidebands and on resonance (represented by different colors/gray levels). Solid lines represent feedback cooling, while the dashed line (for $\Dc=-\om$ only) corresponds to sideband cooling without feedback. (a) $\sss{n_{\m}}$ against coupling $g$ for fixed cavity decay rate $\kappa=\om/2$ (sideband resolved regime). (b) $\sss{n_{\m}}$ against $\kappa$ for $g=\om/10$ (weak coupling regime). Other parameters for both (a) and (b): $Q=5\cdot 10^6$, $\bar{n}=3.5\cdot 10^5$, $h_m=100$, $\eta=1$}
  \label{fig:occ}
\end{figure}

For micro-mechanical systems, however, the sideband-resolved regime \(\kappa<\om\) is typically more relevant. In this regime the picture changes completely. For weak coupling (\(g<\kappa\)) we find two pronounced dips at \emph{both} mechanical sidebands (\(\Dc=\pm\om\)), where \(\sss{n_{\m}}\) is locally minimal and clearly lies below the value on resonance. It is obvious from the figure that cooling works best on the red sideband (\(\Dc=-\om\)), where we have a cumulative effect from passive sideband and feedback cooling (see also \fref{fig:occ}). However, even on the blue sideband (\(\Dc=\om\))---which is commonly associated with heating---we find an appreciable reduction of the mechanical occupation by feedback cooling. As we discussed in the previous section, we can attribute this effect to large optomechanical correlations, which allow for a good read out of the mechanical motion and thus a good feedback performance. If we increase the coupling strength to \(g=0.3\kappa\) we see a peak appearing around the blue sideband (which we attribute to ponderomotive squeezing of the output fields), pushing the occupation number above the value at \(\Dc=0\).
For both regimes we plot graphs for two different detection efficiencies \(\eta=1\) (lossless detection) and \(\eta=8/10\). Clearly, non-unit detection efficiency leads to a noticeable degradation of feedback-cooling performance. Only at the red sideband and in the sideband-resolved regime, where the effect of sideband cooling dominates, the final occupation number is virtually unaffected.

Figure~\ref{fig:occ}(a) shows the mechanical occupation for three detunings \(\Dc=0,\pm\om\) plotted against \(g\). For \(\Dc=-\om\) we show, additionally to the closed-loop feedback solution (red solid line), the solution for sideband cooling (red dashed line). While for \(\Dc=0\) and \(\Dc=-\om\) the occupation number steadily decreases---in the depicted range---for growing \(g\), for \(\Dc=\om\) a clear minimum is visible in the weak coupling regime at \(g\approx \kappa/10\). This minimum lies well below the value for \(\Dc=0\) (but still above the value for the red sideband). This means that there exists a considerably large parameter regime where a detuned operation significantly improves the performance of feedback cooling. Note that all curves rise drastically in the strong-coupling regime, where \(g/\kappa \gtrsim 1\) (not shown in the plot). In \fref{fig:occ}(b) we plot \(\sss{n_{\m}}\) against cavity linewidth \(\kappa\) for constant coupling \(g\). Again we find that feedback on the sidebands works best in the sideband-resolved regime, while in the bad cavity regime working on resonance yields (slightly) better performance. Again, the occupation number is minimized with respect to the homodyne phase \(\phi\) at each point in the plot.

In summary we illustrated that feedback cooling in the resolved-sideband regime is a viable option for cooling the mechanical oscillator into its ground state. It turns out that in this regime driving the system on the blue mechanical sideband yields a lower mechanical occupation number than operating on resonance. As an extension of this protocol we will show in the next section, that a similar setup operating at the same working point can be used to remotely prepare a squeezed mechanical state via time-continuous teleportation.

\subsection{Time-continuous optomechanical teleportation}
\label{sec-2-4}
\label{SEC:optom-teleport}

Time-continuous teleportation is facilitated by what we call a time-continuous Bell measurement \cite{hofer_time-continuous_2013}, as depicted in \fref{fig:setup}(b): The output field \(A\) of an optomechanical system (denoted by \(S\)) is mixed with a second field \(B\) on a beam-splitter. The resulting fields are then sent to two homodyne detection setups which measure the Einstein--Podolsky--Rosen (EPR) quadratures \(x_+=x_a+x_b\) and \(p_-=p_a-p_b\) where \(x_a\), \(x_b\) and \(p_a\), \(p_b\) are the amplitude and phase quadratures of the respective fields. The field \(B\) is prepared in a pure state of Gaussian squeezed white noise, which we denote by \(\ket{M}\), where \(M\in \mathds{C}\) characterizes the squeezing (see \aref{APP:squeezed}); \(|M|\) describes the absolute increase/reduction of the anti-/squeezed quadrature, while \(\arg{(M)}\) determines the squeezing angle. Provided the optomechanical system--field interaction creates entanglement between the mechanical mode and the outgoing light field, the state of \(B\) can be teleported to the mechanical oscillator by applying (instantaneous) feedback proportional to the measurement results of the Bell measurement (\(I_{\pm}\)). This effectively generates dissipative dynamics which drive the mechanical system into a steady state coinciding with the input light state. In \sref{SEC:teleport-generic} we derive the constitutive equations of motion (the conditional master equation and feedback master equation) for a generic system. In this section we will focus on the optomechanical implementation. Technical details are discussed in \sref{SEC:teleport-om-deriv}.

In order to successfully implement continuous teleportation in optomechanical systems we need to appropriately design our measurement setup. To do this we first need a clear picture of the system's dynamics: In the regime \(g\ll \kappa\ll \om\) and for a blue drive with \(\Dc=\om\) the optomechanical interaction is \(H_{\mathrm{om}}\approx g(\cm\cc+\cm^{\dagger}\cc^{\dagger})\). Under the weak-coupling condition (\(g\ll \kappa\)) the cavity follows the mechanical mode adiabatically. We will see that in this regime  we effectively obtain the required entangling interaction between the mirror and the outgoing field. Moreover, the mechanical oscillator resonantly scatters photons into the lower sideband at \(\omega_{\mathrm{c}}=\omega_0-\om\). Spectrally, the photons which are correlated with the mechanical motion are therefore located at this sideband frequency. We thus set up our Bell measurement in the following way: Firstly, we choose the center frequency of the squeezed input light located at the sideband frequency \(\omega_{\mathrm{c}}\). Secondly, we now use heterodyne detection to measure quadratures at the same frequency.

In \sref{SEC:teleport-om-deriv} we show that after adiabatic elimination of the cavity mode and a rotating-wave approximation, the evolution of the conditional mechanical state \(\cond{\rho}^{(\m)}\) in a rotating frame with \(\om\) (neglecting the mechanical frequency shift by the optical-spring effect, see \sref{SEC:teleport-om-deriv}) is described by the SME
\begin{multline}
  \label{eq:14}
  \dd\cond{\rho^{(\m)}}=
  \gamma_-\mathcal{D}[\cm]\cond{\rho^{(\m)}}\dt + \gamma_{+}\mathcal{D}[\cm^{\dagger}]\cond{\rho^{(\m)}}\dt\\
  -\sqrt{\frac{\eta g^2\kappa}{2}}\left\{\mathcal{H}[\ii\mu\eta_+ \cm^{\dagger}]\cond{\rho^{(\m)}}\dW_+ - \mathcal{H}[\nu\eta_+ \cm^{\dagger}]\cond{\rho^{(\m)}} \dW_-\right\},
\end{multline}
where we defined \(\gamma_- =\gamma(\bar{n}+1)+2g^2\mathrm{Re}(\eta_-)\), \(\gamma_+ =\gamma\bar{n}+2g^2\mathrm{Re}(\eta_+)\) and \(\eta_{\pm}=[\kappa/2+\ii(-\Dc\pm\om)]^{-1}\).
The second row describes passive cooling and heating effects via the optomechanical interaction, as has been derived before in the quantum theory of sideband cooling \cite{wilson-rae_theory_2007,marquardt_quantum_2007}. The third row describes the time-continuous Bell measurement, where the squeezing parameter \(M\) is encoded in \(\mu=1-\alpha\) and \(\nu=1+\alpha\), with \(\alpha=(N+M)/(N+M^{*}+1)\) (\aref{APP:squeezed}). The parameter \(N>0\) is connected to \(M\) via \(|M|^2=N(N+1)\). \(\eta\) is the detection efficiency as before. The measured photo-currents of the Bell measurement are
\begin{subequations}
  \label{eq:15}
  \begin{align}
    I_+\dt&= -\ii \sqrt{\eta g^2\kappa/2}\,\mean{\eta_+\cm^{\dagger}-\Hc{}}\dt+\dW_+,\\
    I_-\dt&= \sqrt{\eta g^2\kappa/2}\,\mean{\eta_+\cm^{\dagger}+\Hc{}}\dt+\dW_-,
  \end{align}
\end{subequations}
where \(\dW_{\pm}\) are correlated Wiener increments whose \mbox{(co-)variances} are given by
\begin{subequations}
  \label{eq:16}
  \begin{align}
    w_1\dt&{:=} (\dW_+)^2=[N+1+(M+M^{*})/2]\dt,\\
    w_2\dt&{:=} (\dW_-)^2=[N+1-(M+M^{*})/2]\dt,\\
    w_3\dt&{:=} \dW_+\dW_-=-[\ii(M-M^{*})/2]\dt,
  \end{align}
\end{subequations}
as is shown in \sref{SEC:teleport-generic}. For the choice \(\Dc=\om\) we have \(\eta_+=2/\kappa\) and \(\eta_-=1/(\tfrac{\kappa}{2}+2\ii\om)\). Thus  \(I_{\pm}\) approximately correspond to measurements of the mechanical quadratures \(p_{\m}\) and \(x_{\m}\) respectively. We model the feedback as instantaneous displacements of the mechanical oscillator in phase space, where the feedback strength is proportional to the heterodyne currents \(I_{\pm}(t)\). This is described by Hamiltonian terms \(I_{\pm}(t)F_{\pm}\), where \(F_{\pm}=F_{\pm}^{\dagger}\) are generalized forces. The feedback operators we choose to be \(F_+=-\sqrt{2g^2\kappa}\,\eta_+\,x_{\m}\) and \(F_-=-\sqrt{2g^2\kappa}\,\eta_+\,p_{\m}\), which generate a displacement in \(p_{\m}\) and \(x_{\m}\) respectively. The prefactors of \(F_{\pm}\) (\ie{}, the feedback gain) we chose such that they match the measurement strength of the Bell detection.
The corresponding feedback master equation (in the same rotating frame) can be written as
\begin{multline}
  \label{eq:17}
  \dot{\rho}^{(\m)} =
  \gamma(\bar{n}+1)\mathcal{D}[\cm]\rho^{(\m)} + \gamma \bar{n}\mathcal{D}[\cm^{\dagger}] \rho^{(\m)}\\
  +(4g^2/\kappa)
  \left\{\lambda_{1}(\epsilon)\mathcal{D}[J_1(\epsilon)]+\lambda_2(\epsilon)\mathcal{D}[J_2(\epsilon)]
  \right\}\rho^{(\m)},
\end{multline}
where
\(\epsilon=[1+(4\om/\kappa)^2]^{-1}\) quantifies the suppression of the counter-rotating interaction terms (\ie{}, the optomechanical beam-splitter). This suppression is large (\(\epsilon\ll 1\)) in the sideband-resolved regime where \(\kappa\ll \om\) and small (\(\epsilon\approx 1\)) in the bad-cavity regime (\(\kappa\gtrsim \om\)). The effective Lindblad terms are determined by \(\lambda_i\) and \(J_i\), which are obtained (see \aref{APP:diag}) from the eigenvalue decomposition of the positive matrix
\begin{equation*}
  \Lambda=
  \begin{pmatrix}
    \frac{w_2}{\eta}-\frac{1}{2}(1+\epsilon)&-\frac{w_3}{\eta}+\frac{\ii}{2}(1+\epsilon)\\
    -\frac{w_3}{\eta}-\frac{\ii}{2}(1+\epsilon)&\frac{w_1}{\eta}-\frac{1}{2}(1+\epsilon)
  \end{pmatrix}.
\end{equation*}
For efficient detection (\(\eta=1\)) we obtain \(\lambda_1=(2N+1)+O(\epsilon)\) and \(\lambda_2=O(\epsilon)\), which means that in the sideband-resolved regime the jump operator \(J_2(\epsilon)\) contributes only weakly. In zeroth order in \(\epsilon\) the dominating dissipative dynamics are generated by \(J_1(0) \propto -\ii (2N+1-M-M^{*})x_{\m}+(1+M-M^{*})p_{\m}\). If we define as \(U_{\pi/2}\) the local unitary which effects the canonical transformation \(x_{\m}\rightarrow p_{\m}\), \(p_{\m}\rightarrow -x_{\m}\), we find by comparison with \eqref{eq:94} that \(U_{\pi/2}^{\dagger}J_1(0)U_{\pi/2}\ket{-M}=0\). Taking into account relations \eqref{eq:93} one can easily show that \(U_{\pi/2}{\ket{-M}}=\ket{M}\). This means that \(\ket{M}\bra{M}\) is a dark state of \(D[J_1(0)]\) and thus in the ideal limit of \(\gamma=0\), \(\epsilon=0\), \(\eta=1\) the steady state of \eqref{eq:17} is \(\lim_{t\rightarrow \infty}\rho^{(\m)}(t)=\ket{M}\bra{M}\). Hence, the optical input state is perfectly transferred to the mechanical mode.

Moving away from the ideal case, the protocol's performance is degraded by mechanical decoherence effects (\(\gamma \bar{n}>0\)), counter-rotating terms of the optomechanical interaction which are suppressed by \(\epsilon<1\), and inefficient detection (\(\eta<1\)) which leads to imperfect feedback.
\begin{figure}[htb]
  \begin{minipage}{\columnwidth}\rlap{(a)}{\centerline{\includegraphics[width=\columnwidth]{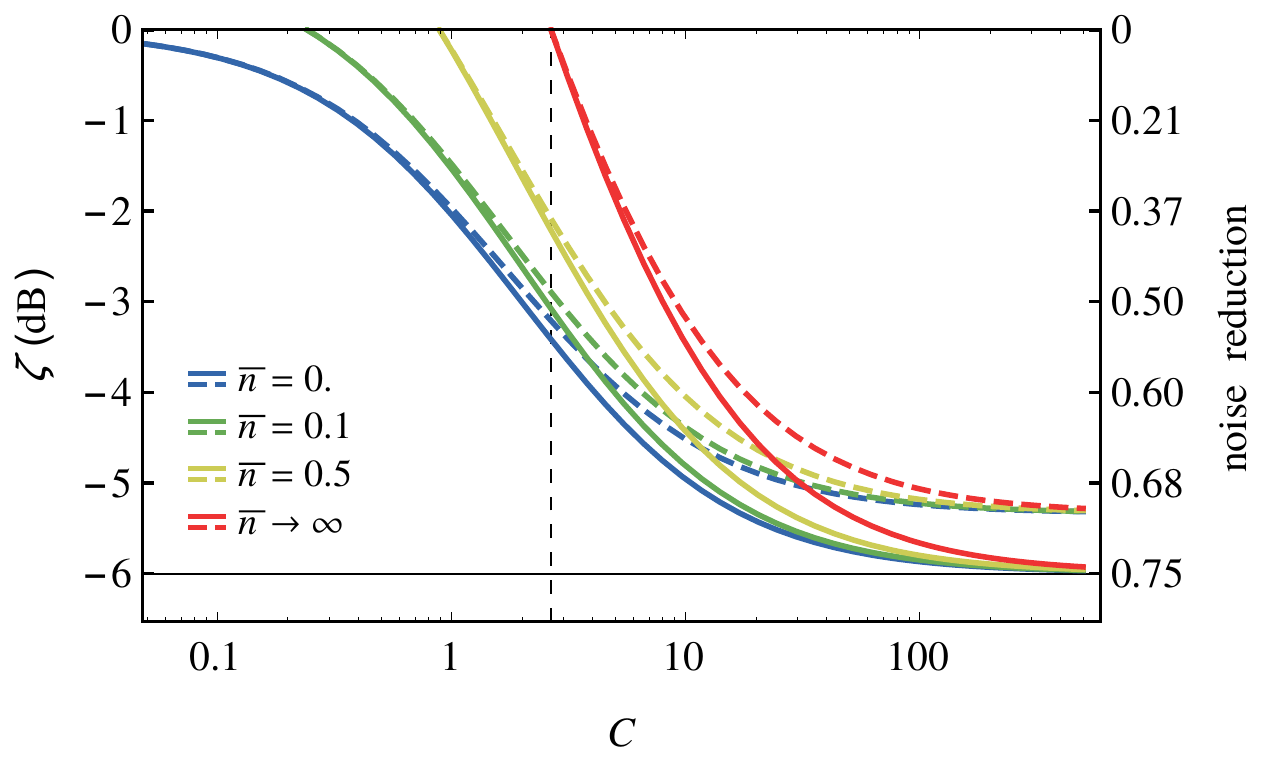}}}\end{minipage}\\
  \begin{minipage}{\columnwidth}\rlap{(b)}{\centerline{\includegraphics[width=\columnwidth]{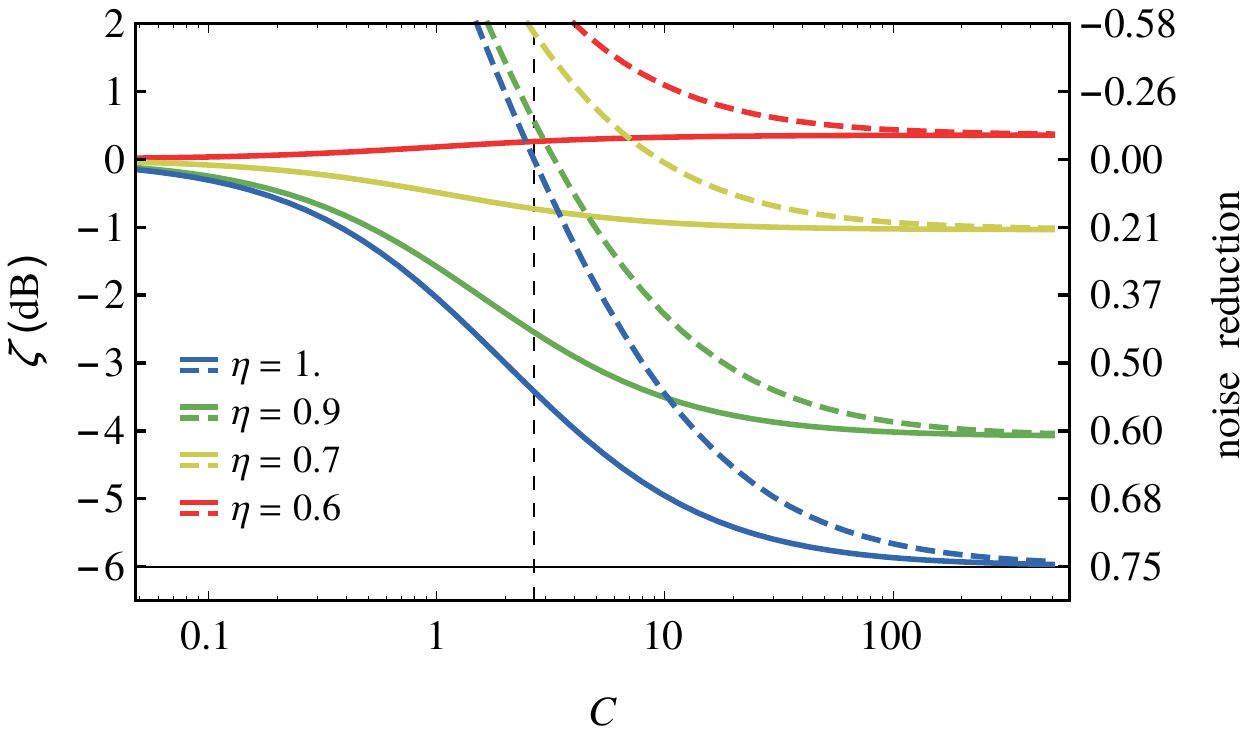}}}\end{minipage}
  \caption[]{(Color online) Mechanical squeezing $\zeta$ against cooperativity $C$: (a) Varying mechanical bath occupation $\bar{n}=0,\ 1/10,\ 1/2,\ \infty$ (represented by different colors/gray levels) and unit detection efficiency $\eta=1$; The solid (dashed) lines represent a sideband resolution of $\kappa/\om= 1/10\ (1)$. (b) Different detection efficienies $\eta=1,\,9/10,\,7/10,\,6/10$ (represented by different colors/gray levels) and $\kappa=\om/10$; Here the solid (dashed) lines represent $\bar{n}=0\ (\infty)$. In both plots the horizontal solid line at $\zeta=-6\mathrm{dB}$ (corresponding to $N\approx 0.56$) shows the squeezing level of the input light and the vertical dashed line the critical cooperativity $C_{\mathrm{crit}}\approx 2.7$.}
  \label{fig:teleport}
\end{figure}
\Fref{fig:teleport} shows the steady-state squeezing \(\zeta\) transmitted to the mechanical mode for different parameters plotted against the optomechanical cooperativity \(C=4g^2/(\bar{n}+1)\gamma\kappa\). In the upper plot we assume perfect detection efficiency \(\eta=1\) and find that in this case there exists a critical value \(C_{\mathrm{crit}}(N)=1/[\sqrt{N(N+1)}]\) determined by the input squeezing \(N\) above which the resulting mechanical state is squeezed for any thermal occupation number \(\bar{n}\). The lower plot clearly shows that this is no longer true if we assume non-unit detection efficiency \(\eta\). We find that below a certain critical value \(\eta_{\mathrm{crit}}(N,\bar{n})\) we can no longer transfer squeezing to the mechanical oscillator, but we rather heat it instead. (This is even true for a zero-temperature mechanical environment, as illustrated in the plot.)
In this general case it can be beneficial to chose a modified feedback gain, \ie{}, use feedback operators \(\tilde{F}_{\pm}=\sigma F_{\pm}\) with \(\sigma\neq 1\). In the parameter regime we consider however the resulting improvement negligible.

\subsection{Time-continuous optomechanical entanglement swapping}
\label{sec-2-5}
\label{SEC:entswap}

We now replace the squeezed field mode with a second optomechanical cavity, as is depicted in \fref{fig:setup}(c). The goal of this scheme is to generate stationary entanglement between the two mechanical subsystems. This is again facilitated by a time-continuous Bell measurement---measuring the output light of both cavities---plus feedback \cite{hofer_time-continuous_2013}.
The implementation is akin to the teleportation protocol presented above: Both cavities are driven on the blue sideband to resonantly enhance the two-mode squeezing interaction, and their output light is sent to the Bell detection setup which operates at the cavity resonance frequency \(\omega_c\). Feeding back the Bell detection results \(I_{\pm}\) corresponding to the \(x_+\) and \(p_-\) quadratures of the optical fields to both mechanical systems dissipatively drives them towards an entangled state. There is a slight complication, however. A single Bell measurement only allows us to separately monitor two of the four variables (\(x_{\m,1},p_{\m,1},x_{\m,2},p_{\m,2}\)) needed to describe the quantum state of the mechanical systems.\footnote{In the language of control theory this means that the complete system is not observable (see for example \cite{wiseman_quantum_2009}).} Combined with the fact that we drive the system on the blue side of the cavity resonance (and thus in an unstable regime) this means that we cannot actively stabilize the system and---depending on the driving strength and sideband resolution---no steady state may exist. To compensate for this we extend the setup by two additional heterodyne detectors, measuring \(x_-\) and \(p_+\) with outcomes \(I'_{\mp}\). The effective measurement strength of this stabilizing measurements with respect to the Bell measurement is set by the transmissivity \(\upsilon\) of the beam-splitter in front of the heterodyne setup (see \fref{fig:setup}). Appropriate feedback of all measurement currents \(I_{\pm}\), \(I'_{\pm}\) (for simplicity labeled \(I_i\), \(i=1,\dots,4\), below) to both mechanical systems finally allows us to stabilize them in an entangled state. Note that this setup effectively realizes two simultaneous Bell measurements of the pairs (\(x_+,p_-\)) and (\(x_-,p_+\)) with detection efficiencies \(\upsilon\) and \(1-\upsilon\) respectively. In the rest of this section the two optomechanical systems are assumed to be identical and all detectors to have the same quantum detection efficiency \(\eta\).

In \sref{SEC:om-swap-deriv} we show that in an adiabatic approximation the conditional state of the two mechanical oscillators \(\rho^{(\m)}\) in a rotating frame can be described by the SME (setting \(\Dc=\om\))
\begin{multline}
  \label{eq:19}
  \dd{\cond{\rho^{(\m)}}}= \epsilon \frac{4g^2}{\kappa} \left( \mathcal{D}[c_{\m,1}]\cond{\rho^{(\m)}}+\mathcal{D}[c_{\m,2}]\cond{\rho^{(\m)}} \right)\dt\\
  + \sum_{i=1}^2 \left\{ \gamma(\bar{n}+1)\mathcal{D}[c_{\m,i}]\cond{\rho^{(\m)}} + \gamma \bar{n}\mathcal{D}[c_{\m,i}^{\dagger}]\cond{\rho^{(\m)}} \right\}\dt\\
  + \frac{2g^2}{\kappa}\sum_{i=1}^4\left(\mathcal{D}[J_i]\cond{\rho^{(\m)}}\dt+\sqrt{\eta} \mathcal{H}[J_i]\cond{\rho^{(\m)}}\dW_i\right),
\end{multline}
where we set \((J_1,J_2)=\sqrt{\upsilon}\bigl(c_{\m,+},\ii c_{\m,-}\bigr)\), \((J_3,J_4)=\sqrt{1-\upsilon}(\ii c_{\m,+},c_{\m,-})\) and \(c_{\m,\pm}=c_{\m,1}\pm c_{\m,2}\).
The Wiener processes \(W_i\) are uncorrelated with unit variance, \ie{}, \(\dW_i\dW_j=\delta_{ij}\dt\), and correspond to the photo-currents
\begin{equation}
  \label{eq:20}
  I_i\dt=\sqrt{4g^2/\kappa}\cond{\mean{J_i+J_i^{\dagger}}}\dt+\dW_i.
\end{equation}

\begin{figure}[tbh]
  \begin{minipage}{\columnwidth}\rlap{(a)}{\centerline{\includegraphics[width=\columnwidth]{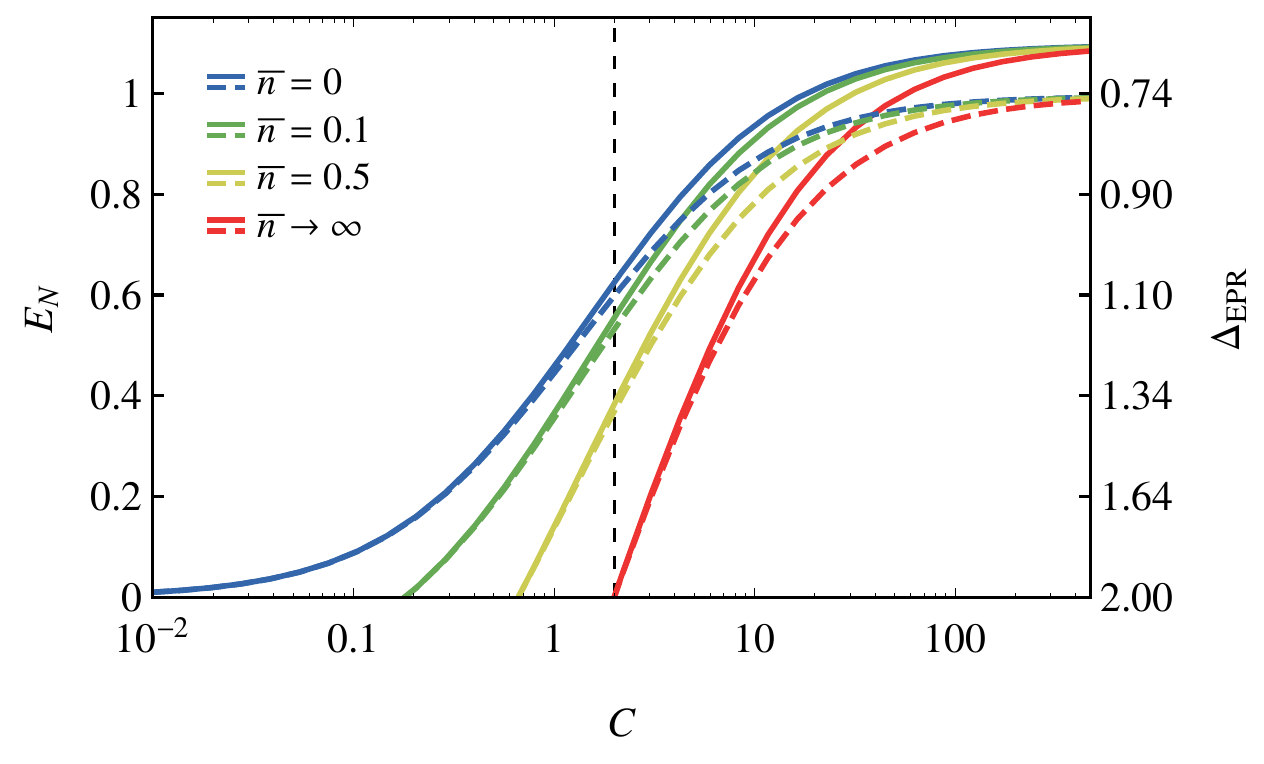}}}\end{minipage}\\
  \begin{minipage}{\columnwidth}\rlap{(b)}{\centerline{\includegraphics[width=\columnwidth]{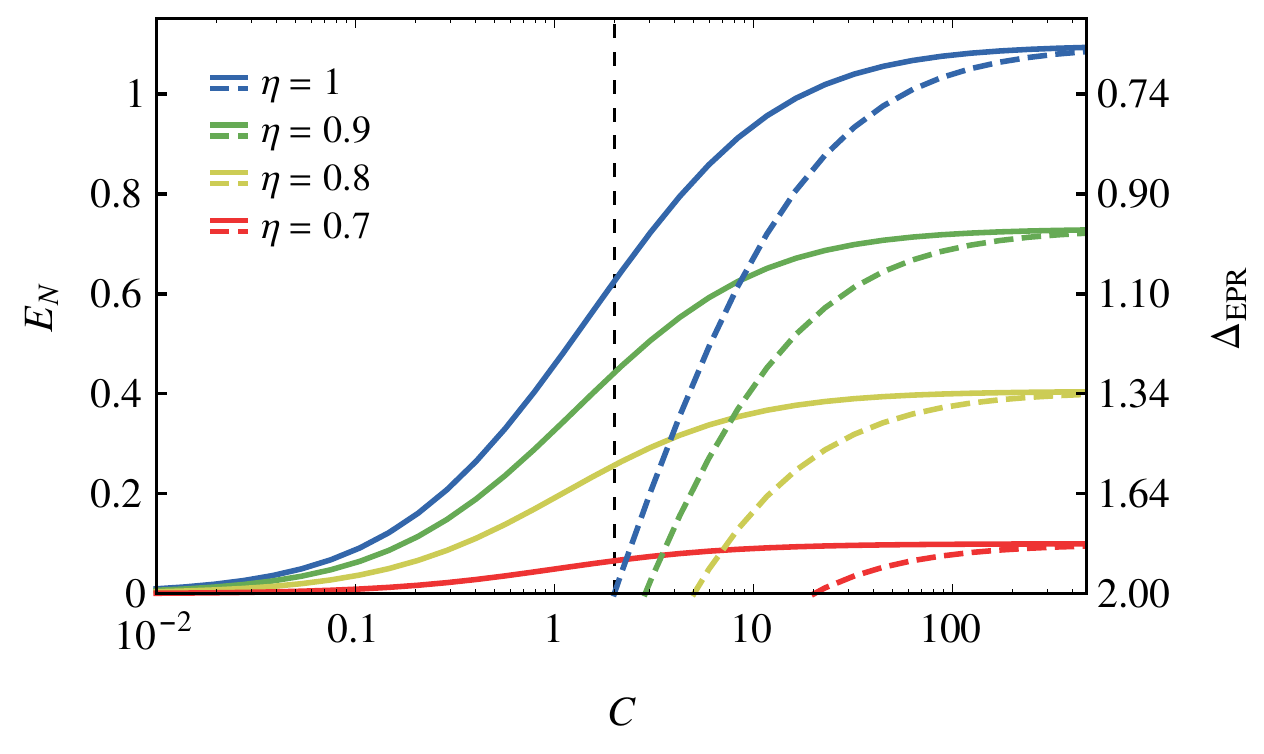}}}\end{minipage}
  \caption[]{(Color online) Two-mode mechanical steady-state entanglement in terms of $\sss{E}_{\mathcal{N}}$ and $\Delta_{\mathrm{EPR}}$ against cooperativity $C$, maximized with respect to feedback gain $\sigma$: (a) Varying mechanical bath occupation $\bar{n}=0,\,1/10,\,1/2,\,\infty$ (represented by different colors/gray levels) for unit detection efficiency $\eta=1$; The solid (dashed) lines represent a sideband resolution of $\kappa/\om= 1\ (1/10)$. (b) Different detection efficienies $\eta=1,\,9/10,\,7/10,\,6/10$ (represented by different colors/gray levels) and $\kappa=\om/10$; Here the solid (dashed) lines represent $\bar{n}=0\ (\infty)$. The black vertical line shows the critical cooperativity $C_{\mathrm{crit}}=2$.}
  \label{fig:entswap}
\end{figure}

The final steady state of this protocol depends on the feedback operators \(F_i=F_i^{\dagger}\) we apply. In ananolgy to the previous section we choose \((F_1,F_2)=\sqrt{\upsilon}\sigma(\ii c_+-\ii c_+^{\dagger},c_-+c_-^{\dagger})\) and \((F_3,F_4)=\sqrt{1-\upsilon}(c_+ + c_+^{\dagger},\ii c_--\ii c_-^{\dagger})\), which can realize independent displacements of all mechanical quadratures. This time we introduced an additional gain parameter \(\sigma\) which we can vary in order to optimize the amount of entanglement in the resulting steady state. With these choices the FME for optomechanical entanglement swapping takes the form
\begin{multline}
  \label{eq:21}
  \cond{\dot{\rho}^{(\m)}}=-\ii[H_{\mathrm{fb}},\cond{\rho^{(\m)}}]+\epsilon \frac{4g^2}{\kappa} \left( \mathcal{D}[c_{\m,1}]+\mathcal{D}[c_{\m,2}] \right)\cond{\rho^{(\m)}}\\
  +\sum_{i=1}^2 \left\{ \gamma(\bar{n}+1)\mathcal{D}[c_{\m,i}]\cond{\rho^{(\m)}} + \gamma \bar{n}\mathcal{D}[c_{\m,i}^{\dagger}]\cond{\rho^{(\m)}} \right\}\\
  + \frac{2g^2}{\kappa}\sum_{i=1}^4 \left\{ \mathcal{D}[J_i-\ii F_i]\cond{\rho^{(\m)}}+\frac{1-\eta}{\eta}\mathcal{D}[F_i] \right\},
\end{multline}
where the dynamics generated by the feedback is described by \(H_{\mathrm{fb}}=\ii [(1+\sigma)\upsilon-1](2g^2/\kappa)(c_{\m,+}^2+c_{\m,-}^2-\Hc{})\). We can now analyze the stability properties of the linear feedback system by evaluating the corresponding Routh--Hurwitz criterion. In the case of no stabilizing feedback (\(\upsilon=1\)) we find that the admissible optomechanical coupling is limited from above by \(4g^2/\kappa<1/(1-\epsilon)\), which only gives an appreciably high limit for values of \(\epsilon\approx 1\) and thus in the bad-cavity regime. The stabilization is caused by the counter-rotating beam-splitter terms \(c_{\m,i}c_{\lm,i}^{\dagger}+\Hc{}\) of the optomechanical Hamiltonian, which cool the mechanical systems. This cooling effect, however, diminishes the amount of generated steady-state entanglement. If we switch on the stabilizing feedback and thus choose \(\upsilon<1\), we can rewrite the RH criterion in the form \([3+(4g^2/\kappa)^{-1}+\epsilon]>4\upsilon>[(1-\epsilon)-(4g^2/\kappa)^{-1}]/\sigma\) (where we assumed \(\epsilon<1\)). These inequalities are tightest in the limit \(\epsilon\rightarrow 0\), \(g^2/\kappa\rightarrow \infty\) where we have \(3>4\upsilon>1/\sigma\). For the rest of this section we choose \(\upsilon=3/4\) which ensures stability of the feedback system for {any} values of \(g^2/\kappa\) and \(\epsilon\)---and consequently the sideband-resolution \(\kappa/\om\)---as long as the feedback gain fulfills \(\sigma>1/3\). In the stable regime and for \(\epsilon=0\), \(\eta=1\) we find a simple analytic expression for the steady-state logarithmic negativity,
\begin{equation}
  \label{eq:22}
  \sss{E}_{\mathcal{N}} = \ln \left(\frac{\frac{1}{2}C(\bar{n}+1)(3 \sigma -1)(4 \upsilon -1)+1}{C (\bar{n}+1)[3 \sigma(\sigma-1)+1]+2 \bar{n}+1}  \right),
\end{equation}
where we again introduced the cooperativity \(C=4g^2/(\bar{n}+1)\gamma\kappa\). Generally we can---for each set of parameters \((C,\bar{n},\epsilon,\upsilon,\eta)\)---maximize the entanglement \(E_{\mathcal{N}}\) with respect to the feedback gain \(\sigma\). In \fref{fig:entswap} we plot the resulting steady-state values in terms of logarithmic negativity \(E_{\mathcal{N}}\) and EPR-variance
\begin{equation}
  \label{eq:23}
  \Delta_{\mathrm{EPR}}=\min_{\phi_1,\phi_2}\left( [\Delta(x_{\m,1}^{\phi_{1}}-x_{\m,2}^{\phi_2})]^2+[\Delta(p_{\m,1}^{\phi_{1}}+p_{\m,2}^{\phi_2})]^2 \right),
\end{equation}
where \(x_{\m,i}^{\phi}=(c_{\m,i}\ee^{-\ii \phi}+c_{\m,i}^{\dagger}\ee^{+\ii \phi})/\sqrt{2}\) and \(p_{\m,i}^{\phi}=x_{\m,i}^{\phi+\pi/2}\) are rotated mechanical quadratures. A Gaussian state is entangled if \(\Delta_{\mathrm{EPR}}<2\) \cite{duan_inseparability_2000,simon_peres-horodecki_2000}. In the first plot we assume a perfect detection efficiency \(\eta=1\) and consider different bath occupation numbers \(\bar{n}\). We again see that there exists a critical cooperativity \(C_{\mathrm{crit}}\) above which we are able to generate entanglement regardless of \(\bar{n}\). From \eqref{eq:22} we can deduce the expression \(C_{\mathrm{crit}}(\upsilon,\sigma)=4/[3\sigma(1+4\upsilon-2\sigma)-(1+4\upsilon)]\). (As is evident from the plot, the \(C_{\mathrm{crit}}\) is independent of \(\epsilon\)). For the parameters used in the plot (taking into account the optimization with respect to \(\sigma\)) we find \(C_{\mathrm{crit}}=2\). Again, counter-rotating terms decrease entanglement but are strongly suppressed by the sideband resolution. In \fref{fig:entswap}(b) we take into account losses and non-unit detection efficiency, \(\eta<1\), which drastically reduces the amount of achieved entanglement. As before we find a critical loss value \(\eta_{\mathrm{crit}}(\bar{n},\upsilon)\) (for the parameters chosen in the plot slightly above 65\%) below which entanglement creation is prohibited.

\section{Derivation of conditional and feedback master equations}
\label{sec-3}
\label{SEC:meq}
We present here a brief (and informal) derivation of the stochastic master equations (SME) and Markovian feedback master equations (FME) we use throughout the paper. A rigorous and complete account of the quantum stochastic formalism and quantum filtering theory can be found in the literature, \eg{}, \cite{hudson_quantum_1984,carmichael_quantum_1993,reynaud_quantum_1997,bouten_introduction_2007,barchielli_quantum_2009,wiseman_quantum_2009}, and a brief summary of the most important relations is given in \aref{APP:qsc}.

\subsection{The homodyne master equation}
\label{sec-3-1}
\label{SEC:homodyne-generic}
We consider a situation similar to \fref{fig:setup}(a), where a system \(S\) couples to the one-dimensional electromagnetic field \(A\), which is initially in vacuum and is subject to homodyne detection. We first assume unit detection efficiency, but will discuss the case of inefficient detection at the end of the section. The system--field coupling is mediated by the Hamiltonian\footnote{Note that both operators \(s\) and \(a(t)\) have a dimension \([s]=[a(t)]=\sqrt{\mathrm{Hz}}\).}
\begin{equation}
  \label{eq:24}
  H_{\mathrm{int}}=\ii [s\,a^{\dagger}(t)-s^{\dagger}a(t)],
\end{equation}
where \(s\) is a system operator (\eg{}, a cavity creation or destruction operator), and the light field is described (in an interaction picture at a central frequency \(\omega_0\)) by \(a(t)=\int \dd{\omega}\, a_0(\omega)\, \ee^{-\ii(\omega-\omega_0)t}\) \cite{gardiner_input_1985}, where \(a_0(\omega)\) is the (Schr\"{o}dinger) annihilation operator of the field mode at \(\omega\). In a Markov approximation the field operators are \(\delta\)-correlated, and fulfill the commutation relations
\begin{equation}
  \label{eq:25}
  [a(t),a^{\dagger}(t')]=\delta(t-t').
\end{equation}
Under this approximation we can introduce the \Ito{} increments \(\dd{A}\), \(\dd{A}^{\dagger}\), which (formally) obey \(\dd{A}(t)=a(t)\dt\), etc., and the vacuum multiplication table given in \aref{APP:qsc}.

The Schr\"{o}dinger equation for the full system (\(S+A\)) initially in the state \(\ket{\phi_0}=\ket{\psi_0}_S\ket{\mathrm{vac}}_A\) can be written in \Ito{}-form as
\begin{equation}
  \label{eq:26}
  \dd{\ket{\phi(t)}} = \left\{  -\ii H_{\mathrm{eff}}\dt+s \left[ \dd{A^{\dagger}}(t)+\dd{A}(t) \right] \right\}\ket{\phi(t)},
\end{equation}
where we used the fact that \(\dd{A}(t)\ket{\phi(t)}=\dd{A}(t)\ket{\phi_0}=\dd{A}(t)\ket{\mathrm{vac}}=0\) \cite{gardiner_quantum_2004}.
A homodyne measurement of an electromagnetic quadrature \(x=a+a^{\dagger}\) with a result \(I_x\) effectively projects the state of the light field onto the eigenstate \(\ket{I_x}\) of \(x\), where \(x\ket{I_x}=I_x\ket{I_x}\) \cite{goetsch_linear_1994}. Projecting \eqref{eq:26} onto \(\ket{I_x}\) leads to the linear stochastic Schr\"{o}dinger equation (SSE) \cite{carmichael_open_1993}
\begin{equation}
  \label{eq:27}
  \dd{\ket{\cond{\tilde{\psi}}(t)}} =\left[-\ii H_{\mathrm{eff}}\dt +\,s\, I_x(t)\dt\right]\ket{\cond{\psi}(t)},
\end{equation}
with the forward pointing \Ito{}-increment \(\dd{\ket{\tilde{\psi}_{\mathrm{c}}(t)}}={\ket{\tilde{\cond{\psi}}(t+\dt)}}-\ket{\cond{\psi}(t)}\). \(\ket{\cond{\tilde{\psi}}}\) is the unnormalized system state which is conditioned on the homodyne photo-current \(I_{x}\). \(I_x\) is \(\delta\)-correlated, \ie{}, \(\mean{I_x(t)I_x(s)}=\delta(t-s)\), and its probability distribution is given (for a fixed time \(t\)) by \(\Upsilon_t(I_x)=|\scp{I_x}{\phi(t+\dt)}|^2\) \cite{goetsch_linear_1994,wiseman_quantum_2009}. Using this, one can show that \(I_x\) can be written as \cite{gardiner_quantum_2004,wiseman_quantum_2009}
\begin{equation}
  \label{eq:28}
  I_x(t)\dt=\cond{\mean{s+s^{\dagger}}}(t)\dt+\dW(t),
\end{equation}
where \(W\) is a classical Wiener process with \(\dW(t)^2=\dt\) and \(\mathbb{E}[\dW(t)]=0\). Here the conditional expectation value should be read as \(\cond{\mean{A}}(t)=\bra{\cond{\psi}(t)}A\ket{\cond{\psi}(t)}\). We can introduce the classical stochastic process \(\widetilde{X}\) defined by \(\dd{\widetilde{X}}(t)=I_x(t)\dt\), which is statistically equivalent to \(\dd{A}(t)+\dd{A}(t)^{\dagger}\). This is due to non-demolition properties of the measurement operator, see \cite{bouten_introduction_2007}. It obeys \(\dd{\widetilde{X}}(t)^2=\dt\).

The corresponding equation of motion for the unnormalized conditional state \(\cond{\tilde{\rho}}=\ket{\tilde{\cond{\psi}}}\bra{\tilde{\cond{\psi}}}\) can be deduced from \eqref{eq:27},
\begin{align}
  \dd\cond{\tilde{\rho}}(t)&=\cond{\tilde{\rho}}(t+\dt)-\cond{\rho}(t) \notag\\
  \label{eq:29}
                           &= \mathcal{L}\cond{\rho}(t)\dt+\left[ s \cond{\rho}(t) +\cond{\rho}(t) s^{\dagger} \right]\dd{\widetilde{X}}(t)
\end{align}
where we used \Ito{} calculus as presented in \aref{APP:qsc}. Here the Liouvillian \(\mathcal{L}\) is given by
\begin{equation}
  \label{eq:30}
  \mathcal{L}{\rho} = -\ii \left( \Heff {\rho} - {\rho}\Heff^{\dagger} \right) + s{\rho} s^{\dagger}.
\end{equation}
Equation \eqref{eq:29} is the quantum analog to the classical Zakai equation \cite{xiong_introduction_2008}. Note that although the Liouvillian is trace-preserving (\ie{}, \(\tr{\mathcal{L}\rho}=0\)), the second term in \eqref{eq:29} does not possess this property. The equation for the normalized state \(\cond{\rho}(t)=\cond{\tilde{\rho}}(t)/\tr{\cond{\tilde{\rho}}(t)}\) is then found by noting that
\begin{equation}
  \label{eq:31}
  \tr{\cond{\tilde{\rho}}(t+\dt)}=1+\cond{\mean{s+s^{\dagger}}}(t)\dd{\widetilde{X}}(t),
\end{equation}
where now \(\cond{\mean{A}}=\tr{\cond{\rho}(t)A}\) and we used \(\tr{\cond{\rho}(t)}=1\). Thus we find
\begin{equation}
  \label{eq:32}
  \tr{\cond{\tilde{\rho}}(t+\dt)}^{-1}= 1 - \cond{\mean{s+s^{\dagger}}}(t)\dd{\widetilde{X}}(t) +  [\cond{\mean{s+s^{\dagger}}}(t)]^2\dt,
\end{equation}
which is obtained by expanding \(\tr{\cond{\tilde{\rho}}(t+\dt)}^{-1}\) to \emph{second order} in \(\dd{\widetilde{X}}\) (which leads to a first-order expansion in \(\dt\)). Using \Ito{} multiplication rules this leads to
\begin{align}
  \dd{\cond{\rho}}(t)&=\frac{\cond{\tilde{\rho}}(t+\dt)}{\tr{\cond{\tilde{\rho}}(t+\dt)}} - \cond{\rho}(t)\notag\\
  \label{eq:34}
                     &=\mathcal{L}\cond{\rho}(t)\dt+\mathcal{H}[s]\cond{\rho}(t)\dW(t),
\end{align}
which is the desired result \cite{gardiner_quantum_2004,wiseman_quantum_2009}. The (nonlinear) measurement term is given by
\begin{equation}
  \label{eq:35}
  \mathcal{H}[s]\cond{\rho}=\bigl(s-\tr{\cond{\rho}s}\bigr)\cond{\rho}+\cond{\rho}\bigl(s-\tr{\cond{\rho}s}\bigr)^{\dagger}.
\end{equation}
It is clear from the derivation that under made assumptions the SSE \eqref{eq:27} is equivalent to the SME \eqref{eq:34}. The stochastic master equation is more general, however, as it can accommodate for additional, unobserved decay channels (such as photon losses/inefficient detection or coupling of a mechanical oscillator to a heat bath), as well as for mixed initial states.

We can generalize the homodyne master equation in several ways: Above we assumed a measurement of a specific light quadrature \(x=a+a^{\dagger}\). To measure a rotated quadrature \(x^{\phi}=a\ee^{-\ii \phi}+a^{\dagger}\ee^{+\ii \phi}\) we have to make the replacement \(s\rightarrow \ee^{\ii \phi} s\), which simply follows from replacing \(a\rightarrow \ee^{-\ii \phi} a\). We can as easily obtain the SME corresponding to heterodyne detection at a LO frequency \(\omega_{\mathrm{lo}}\neq \omega_0\) by substituting \(s\rightarrow \ee^{\ii \Delta_{\mathrm{lo}} t} s\), where \(\Delta_{\mathrm{lo}}=\omega_{\mathrm{lo}}-\omega_0\).
Below we will discuss the situation where we split up the field with a beam-splitter (with a splitting ration \(\eta:1-\eta\)) and perform two simultaneous homodyne measurements on its outputs. The measured modes \(A'\), \(B\) after the beam-splitter [see \fref{fig:setup}(a)] are related to \(A\) before the beam-splitter via \cite{gough_series_2009} \(a(t)=\sqrt{\eta}\,a'(t)+\sqrt{1-\eta}\,b(t)\), where \(A'\) and \(B\) are both initially in vacuum and are uncorrelated such that \(\dd{A}(t)\dd{B}^{\dagger}(t)=0\), etc. Plugging this relation into \eqref{eq:26} and projecting onto the quadratures \(\ee^{-\ii \phi_1}a' + \ee^{\ii \phi_1}a'^{\dagger}\), \(\ee^{-\ii \phi_2}b + \ee^{\ii \phi_2}b^{\dagger}\) one can repeat above steps and find
\begin{multline}
  \label{eq:36}
  \dd{\cond{\rho}} = \mathcal{L}\cond{\rho}\dt+\sqrt{\eta}\mathcal{H}[\ee^{\ii \phi_1}s]\cond{\rho}\dW_1\\
  +\sqrt{1-\eta}\mathcal{H}[\ee^{\ii \phi_2}s]\cond{\rho}\dW_2,
\end{multline}
with uncorrelated Wiener processes \(W_i\), \ie{}, \(\dW_i\dW_j=\delta_{ij}\). To model photon losses or inefficient photo-detectors, we average over, say, the second measurement process, and thus discard all information obtained from it. Due to the fact that \(\ev{\dW_2}=0\), the equation of motion for the resulting conditional state---which is now conditioned on the measurement results of the first channel only---is obtained by dropping the last term in \eqref{eq:36}. The beam-splitter transmissivity is then identified with the efficiency of the photo-detection. Formally we can obtain the same result from \eqref{eq:34} by replacing \(s \rightarrow \sqrt{\eta}s\) in the measurement term, while keeping the Liouvillian unchanged.

\subsection{Time-continuous teleportation}
\label{sec-3-2}
\label{SEC:teleport-generic}

\subsubsection{Conditional master equation}
\label{sec-3-2-1}
We now extend the derivation of the homodyne SME presented in the previous section to the Bell-measurement setup depicted in \fref{fig:setup}(b). Again, a one-dimensional field mode \(A\) [described by \(a(t)\)] couples to a system \(S\) via \eqref{eq:24}. \(A\) is assumed to be in the vacuum state. A second field mode, \(B\) [with a field operator \(b(t)\)] is prepared in a pure squeezed state, parametrized by \(M\in \mathbb{C}\), which we simply denote by \(\ket{M}\). \(M\) describes the degree and angle of squeezing, as described in \aref{APP:squeezed}. The \Ito{} multiplication table for \(B\) is thus given by
\begin{center}
  \begin{tabular}{c|ccc}
    \(\times\) & \(\dd{B}\) & \(\dd{B}^\dagger\) & \(\dt\)\\
    \hline
    \(\dd{B}\) & \(M\dt\) & \((N+1)\dt\) & 0\\
    \(\dd{B}^\dagger\) & \(N\dt\) & \(M^{*}\dt\) & 0\\
    \(\dt\) & 0 & 0 & 0\\
  \end{tabular}
\end{center}
with the condition  \(|M|^2=N(N+1)\). \(A\) and \(B\) are combined on a balanced beam splitter, whose output is sent to two homodyne detection setups, which are set up such that they measure the EPR operators \(x_+=(a+a^{\dagger}+b+b^{\dagger})/\sqrt{2}\) and \(p_-=\ii(a-a^{\dagger}-b+b^{\dagger})/\sqrt{2}\). We call the continuous measurement of the two quadratures \(x_+\) and \(p_-\) a time-continuous Bell measurement \cite{hofer_time-continuous_2013}. To find the corresponding SME, describing the state of \(S\) conditioned on measurements of \(x_+\) and \(p_-\), we apply the same reasoning as in the previous section. We start from the Schr\"{o}dinger equation \eqref{eq:26} but now choose the initial condition \(\ket{\phi_0}=\ket{\psi_0}_S\ket{\mathrm{vac}}_A\ket{M}_B\). Using the eigenvalue equation for squeezed states \eqref{eq:94}, written as \([\dd{B}(t)-\alpha\dd{B}(t)]\ket{M}=0\) with \(\alpha=(N+M)/(N+M^{*}+1)\), and again \(\dd{A}(t)\ket{\mathrm{vac}}=0\), we can write
\begin{align}
  \label{eq:37}
  \dd{\ket{\phi}} &= \left\{-\ii H_{\mathrm{eff}}\dt+s\left[ \dd{A^{\dagger}}-\alpha\dd{A}+\dd{B}-\alpha\dd{B}^{\dagger}\right]\right\}\ket{\phi}\notag\\
                  &=\left\{-\ii H_{\mathrm{eff}}\dt+\sqrt{1/2}\,s\left[ \mu\dd{X_+}+\ii\nu\dd{P_-} \right] \right\}\ket{\phi},
\end{align}
where \(\mu=1-\alpha\), \(\nu=1+\alpha\) and \(\dd{X_+}(t)=x_+(t)\dt\), \(\dd{P_-}(t)=p_-(t)\dt\). Going from the first to the second line we used the fact that \(a^{\dagger}+b=(x_+ + \ii p_-)/\sqrt{2}\). We emphasize that \(x_+\) and \(p_-\) commute and can be measured simultaneously. We can thus directly project equation \eqref{eq:37} onto the EPR state \(\ket{I_+I_-}_{AB}\) defined by \(x_+\ket{I_+I_-}_{AB}=I_+\ket{I_+I_-}_{AB}\) and \(p_-\ket{I_+I_-}_{AB}=I_-\ket{I_+I_-}_{AB}\). This yields the linear stochastic Schr\"{o}dinger equation
\begin{equation}
  \label{eq:38}
  \dd{\ket{\tilde{\cond{\psi}}}} =\left\{-\ii H_{\mathrm{eff}}\dt +\sqrt{1/2}\,s[\mu\dd{\widetilde{X}}_++\ii \nu\dd{\widetilde{P}}_-]\right\}\ket{\cond{\psi}},
\end{equation}
where \(\dd{\widetilde{X}}_+(t)=I_+(t)\dt\) and \(\dd{\widetilde{P}}_-(t)=I_-(t)\dt\) are again classical processes which possess the same statistical properties as their quantum counterparts. The photo-currents \(I_{\pm}\) (analogous to the previous section) can be written as
\begin{subequations}
  \label{eq:39}
  \begin{align}
    I_+\dt&=\sqrt{1/2}\cond{\mean{s+s^{\dagger}}}\dt+\dW_+,\\
    I_-\dt&=\ii\sqrt{1/2}\cond{\mean{s-s^{\dagger}}}\dt+\dW_-,
  \end{align}
\end{subequations}
with Wiener increments \(\dW_{\pm}\). Comparison to the output of a single homodyne setup \eqref{eq:28} shows that \(I_{\pm}\) correspond to two simultaneous homodyne measurements with half efficiency. The (co-)variances of \(\dW_\pm\) are given in equations~\eqref{eq:16} and directly follow from the definition of \(\dd{X_+}\), \(\dd{P_-}\) and the multiplication tables for \(\dd{A}\) and \(\dd{B}\). We now repeat the procedure from the previous section which is now more involved due to fact that we have to deal with two correlated random processes. It is convenient to introduce the complex process \(\dd{Y}(t)=\mu\dd{\widetilde{X}_+}(t)+\ii \nu \dd{\widetilde{P}_-}(t)\), which obeys \(\dd{Y}^2=2\zeta\dt{:=}-2N/M^{*}\dt\) and \(|\dd{Y}|^{2}=2\dt\). We can then write the Zakai equation corresponding to \eqref{eq:38} as
\begin{equation}
  \label{eq:40}
  \dd{\tilde{\cond{\rho}}}=\mathcal{L}\cond{\rho}\dt+\sqrt{1/2}\left(s\cond{\rho}\dd{Y}+\cond{\rho} s^{\dagger}\dd{Y}^{*}\right),
\end{equation}
with \(\mathcal{L}\) defined in \eqref{eq:30}. To normalize this equation we first calculate
\begin{equation}
  \label{eq:41}
  \tr{\tilde{\cond{\rho}}(t+\dt)}=1+\sqrt{1/2} \bigl[ \cond{\mean{s}}(t)\dd{Y}(t)+\Hc{} \bigr],
\end{equation}
which we use to obtain (by expanding to second order in \(\dd{Y}\))
\begin{multline}
  \label{eq:42}
  \tr{\tilde{\cond{\rho}}(t+\dt)}^{-1}=1-\sqrt{1/2} \bigl[ \cond{\mean{s}}(t)\dd{Y}(t)+\Hc{} \bigr]\\
  +(1/2) \left\{ [\cond{\mean{s}}(t)]^2\zeta + [\cond{\mean{s^{\dagger}}}(t)]^{2}\zeta^{*} +4|\cond{\mean{s}}(t)|^2 \right\}\dt.
\end{multline}
Combining this with \eqref{eq:40} we find after some algebra
\begin{equation}
  \label{eq:43}
  \begin{aligned}
    \dd{\cond{\rho}}&=\mathcal{L}\cond{\rho}\dt+\sqrt{1/2}\left[\mu\bigl(s-\cond{\mean{s}}\bigr)\cond{\rho}+\Hc{}\right]\cond{\rho}\dW_+\\
    &\quad+\sqrt{1/2}\left[\ii\nu\bigl(s-\cond{\mean{s}}\bigr)\cond{\rho}+\Hc{}\right]\cond{\rho}\dW_-\\
    &=\mathcal{L}\cond{\rho}\dt+\sqrt{1/2} \left\{ \mathcal{H}[\mu s]\cond{\rho}\dW_+ +\mathcal{H}[\ii \nu s]\cond{\rho}\dW_- \right\}.
  \end{aligned}
\end{equation}
It can easily be checked that this is a equation of the form \eqref{eq:73} and thus a valid Belavkin equation \cite{belavkin_quantum_1992}.

To conclude this section let us briefly discuss, as a slight variation of above setup, the situation where instead of the squeezed state \(\ket{M}_B\) we use a displaced squeezed state \(\ket{M,\beta}_{B}=D(\beta)\ket{M}_{B}\) (see \aref{APP:squeezed}) as an initial state of mode \(B\), and thus as an input state for teleportation. Transforming the Schr\"{o}dinger equation \eqref{eq:26} into a displaced frame with \(D(\beta)\) shows that the structure of the SSE \eqref{eq:27} and the SME \eqref{eq:43} remains unchanged, if the measurement processes are replaced by appropriately displaced versions, \(\dd{\widetilde{X}}_+ \rightarrow \dd{\widetilde{X}}_+ - \sqrt{1/2}(\beta+\beta^{*})\dt\) and \(\dd{\widetilde{P}}_- \rightarrow \dd{\widetilde{P}}_- +\ii \sqrt{1/2}(\beta-\beta^{*})\dt\). Consequently, the same transformation has to be applied to the currents \(I_{\pm}\) in equations~\eqref{eq:39}.
A more rigorous derivation of the results in this section is presented in \cite{dabrowska_belavkin_2014}.

\subsubsection{Feedback master equation}
\label{sec-3-2-2}
We follow \cite{wiseman_quantum_1993-1} to apply Markovian (\ie{}, instantaneous) feedback proportional to the homodyne currents \(I_{\pm}\) to the system \(S\). This procedure consists of three steps: Converting the \Ito{}-equation for the conditional state into Stratonovich form \cite{gardiner_quantum_2004}, adding a feedback term, and converting back to \Ito{}-form in order to average over all possible measurement trajectories and to obtain an unconditional master equation. The feedback we model as generalized forces \(F_{\pm}=F_{\pm}^{\dagger}\) in the form of additional Hamiltonian terms which we write as \(\mathcal{K}_{\pm}\rho I_{\pm}=-\ii [F_{\pm}I_{\pm},\rho]\).

We start by rewriting the SME \eqref{eq:43} in terms of the complex Wiener increment \(\dd{W_y}=\mu\dd{W}_++\ii \nu \dd{W}_-\) (this time including the detector efficiency \(\eta\) as discussed in \sref{SEC:homodyne-generic}),
\begin{equation}
  \label{eq:44}
  \dd{\cond{\rho}} = \mathcal{L}\cond{\rho}\dt+\sqrt{\eta/2}\left(\mathcal{G}\cond{\rho}\dd{W_y}+\mathcal{G}^{\dagger}\cond{\rho}\dd{W_y}^{*}\right),
\end{equation}
where we defined \(\mathcal{G}\rho=(s-\tr{s\rho})\rho\) and \(\mathcal{G}^{\dagger}\rho=(\mathcal{G}\rho)^{\dagger}\). The Stratonovich form of this equation is given by
\begin{equation}
  \label{eq:45}
  \begin{split}
    (\mathbf{S})\quad\dd{\cond{\rho}} & =
    \mathcal{L}\cond{\rho}\dt+\sqrt{\frac{\eta}{2}}\left(\mathcal{G}\cond{\rho}\dd{W_y}+\mathcal{G}^{\dagger}\cond{\rho}\dd{W_y}^{*}\right) \\
    &\qquad-\frac{\eta}{4}\left(\mathcal{G}\cond{\rho}\dd{W_y}+\mathcal{G}^{\dagger}\cond{\rho}\dd{W_y}^{*}\right)^2\\
    & =
    \mathcal{\bar{L}}\cond{\rho}\dt+\sqrt{\frac{\eta}{2}}\left(\mathcal{G}\cond{\rho}\dd{W_y}+\mathcal{G}^{\dagger}\cond{\rho}\dd{W_y}^{*}\right),
  \end{split}
\end{equation}
with the definition \(\mathcal{\bar{L}}=\mathcal{L}-\tfrac{\eta}{2}[\zeta \mathcal{G}^2+\zeta^{*}(\mathcal{G}^{\dagger})^2+2 \mathcal{G^{\dagger}G}]\). Here we used the fact that \(\mathcal{G^{\dagger}G}=\mathcal{GG^{\dagger}}\). Adding feedback terms \([\cond{\dot{\rho}}]_{\mathrm{fb}}=\sqrt{1/2\eta}[\mathcal{K}_+ I_+ + \mathcal{K}_- I_-]\) and converting back to \Ito{} form yields
\begin{multline}
  \label{eq:46}
  \dd{\cond{\rho}}=\mathcal{L}\cond{\rho}+(1/4\eta)\bigl( \mathcal{K}_+\dd{\widetilde{X}}_++\mathcal{K}_-\dd{\widetilde{P}}_- \bigr)^2\cond{\rho}\\
  +(1/2) \bigl( \mathcal{K}_+\dd{\widetilde{X}}_++\mathcal{K}_-\dd{\widetilde{P}}_- \bigr) \left(\mathcal{G}\dd{W_y}+\mathcal{G}^{\dagger}\dd{W_y}^{*}\right)\cond{\rho}\\
  +\sqrt{1/2\eta}\bigl( \mathcal{K}_+\dd{\widetilde{X}}_++\mathcal{K}_-\dd{\widetilde{P}}_- \bigr)\\
  +\sqrt{\eta/2}\left(\mathcal{G}\dd{W_y}+\mathcal{G}^{\dagger}\dd{W_y}^{*}\right)\cond{\rho},
\end{multline}
where we chose an ordering \(\mathcal{KG}\) to get a trace preserving master equation \cite{wiseman_quantum_1993-1}.
We can now average over all possible measurement trajectories (\ie{}, over all classical processes \(\widetilde{X}_+\), \(\widetilde{P}_-\), \(W_y\)) to obtain an unconditional (deterministic) master equation. A longish calculation leads to
\begin{multline}
  \label{eq:47}
  \dot{\rho}=\mathcal{L}\rho+\frac{1}{2}\left\{ \frac{w_1-w_3}{\eta}\mathcal{D}[F_+]-\ii [F_+,s\rho+\rho s^{\dagger}] \right\}\\
  +\frac{1}{2}\left\{ \frac{w_2-w_3}{\eta}\mathcal{D}[F_-]-\ii [F_-,(\ii s)\rho+\rho (\ii s)^{\dagger}] \right\}\\
  +\frac{w_3}{2\eta}\mathcal{D}[F_++F_-],
\end{multline}
where we used the fact that \(\tfrac{1}{2}(\mathcal{K}_{\pm})^2= \mathcal{D}[F_{\pm}]\) and \(\tfrac{1}{2}(\mathcal{K_+K_-}+\mathcal{K_-K_+})=\mathcal{D}[F_++F_-]-\mathcal{D}[F_-]-\mathcal{D}[F_+]\). \(w_i\) are the (co-)variances of \(\dW_+\), \(\dW_-\) and are given by \eqref{eq:16}. Using the identity \(\mathcal{D}[s+\ii F_{\pm}]\rho=\mathcal{D}[s]\rho+\mathcal{D}[F_{\pm}]\rho+\ii [F_{\pm},s\rho+\rho s^{\dagger}]+\tfrac{\ii}{2}[\rho,F_{\pm}s+s^{\dagger}F_{\pm}]\) this can be written in the more familiar form
\begin{multline}
  \label{eq:48}
  \dot{\rho} = -\ii \left[ H+(1/4)\left\{(F_{+}+\ii F_{-})s+s^{\dagger} (F_+ -\ii F_{-})\right\},\rho \right] \\ + (1/2)\big\{\mathcal{D}[s-\ii F_+]\rho +\mathcal{D}[s- F_{-}]\rho + \frac{w_3}{\eta} \mathcal{D}[F_++F_{-}]\rho\\ +\left(\frac{w_1-w_3}{\eta}-1\right)\mathcal{D}[F_+]\rho+\left(\frac{w_2-w_3}{\eta}-1\right)\mathcal{D}[F_{-}]\rho\big\}.
\end{multline}
If we consider again the situation where we use a displace squeezed state \(\ket{M,\beta}_B\) as input, we make the replacements \(\dd{\widetilde{X}}_+ \rightarrow \dd{\widetilde{X}}_+ - \sqrt{1/2}(\beta+\beta^{*})\dt\) and \(\dd{\widetilde{P}}_- \rightarrow \dd{\widetilde{P}}_- + \ii \sqrt{1/2}(\beta-\beta^{*})\dt\) in equation \eqref{eq:46}. This only changes the third line as all products \(\dd{\widetilde{X}}^2\), \(\dd{\widetilde{X}}\dd{W_y}\), etc.~are unaffected. After taking the classical average this yields an additional Hamiltonian term \(H_{\mathrm{coh}}= \sqrt{2}[\Re(\beta)F_+ - \Im(\beta) F_-]\) which has to be incorporated into \(\mathcal{L}\) in the FME \eqref{eq:47}.

Note that equation \eqref{eq:48} is not necessarily a Lindblad equation, as the prefactors to the operators \(\mathcal{D}\) may in general be negative. It can easily be brought into Lindblad form, however, see \aref{APP:diag}.

\subsection{Time-continuous entanglement swapping}
\label{sec-3-3}
\label{SEC:swap-generic}

\subsubsection{Conditional master equation}
\label{sec-3-3-1}
Consider the setup depicted in \ref{fig:setup}(c): Two systems \(S_1\) and \(S_2\) couple to field modes \(A\) and \(B\) (described by field operators \(a\) and \(b\), both in vacuum) via interaction Hamiltonians analog to \eqref{eq:24}. \(A\) and \(B\) are combined on a 50:50 beam-splitter to form the combinations \(a\pm b\) in the outputs. These outputs are sent to a pair of beam-splitters (with an uneven splitting ratio \({\upsilon:1-\upsilon}\)) and subsequently to a total of four homodyne setups. If we label the modes incident on the homodyne detectors as \(C_i\) (described by field operators \(c_i\)) for \(i=1\dots4\) [see \fref{fig:setup}(d)], we find the following relations to modes \(A\) and \(B\),
\begin{align}
  \label{eq:49}
  a&=\sqrt{\upsilon/2}(c_1+c_2)-\sqrt{(1-\upsilon)/2}(c_3+c_4),\\
  b&=\sqrt{\upsilon/2}(c_1-c_2)-\sqrt{(1-\upsilon)/2}(c_3-c_4).
\end{align}
We now choose the LO phases of the four homodyne setups such that they measure \(x_+{=}c_1+c_1^{\dagger}\), \(p_-{=}-\ii(c_2-c_2^{\dagger})\), \(x_-{=}c_4+c_4^{\dagger}\) and \(p_+{=}-\ii(c_3-c_3^{\dagger})\).
These four measurements allow us to simultaneously monitor both quadratures of both systems (although with imperfect precision). The measurement of \(x_+\) and \(p_-\), which we choose to have a relative strength \(\upsilon\) set by the beam-splitting ratio, realize a Bell measurement as before, while the measurement of the conjugate quadratures \(x_-\) and \(p_+\), with a strength \(1-\upsilon\), we will need for stabilization of \(S_1\) and \(S_2\).

To derive the SME we apply the same logic as before. We start from the Schr\"{o}dinger equation for the full system (\(S_1+S_2+\text{field modes}\)),
\begin{equation}
  \label{eq:50}
  \dd{\ket{\phi}} = \left[-\ii H_{\mathrm{eff}}\dt+s_1 \dd{A^{\dagger}}+s_2 \dd{B^{\dagger}}\right]\ket{\phi},
\end{equation}
with an initial state \(\ket{\phi}=\ket{\psi (0)}_{S_1S_2}\ket{\mathrm{vac}}_{\mathrm{field}}\) and an effective Hamiltonian \(H_{\mathrm{eff}} = H_{\mathrm{sys}}^\mathrm{(1)}+H_{\mathrm{sys}}^\mathrm{(2)}-\frac{\ii}{2} \sum_{i=1,2}s_i^{\dagger}s_i\). We then rewrite this in terms of \(\dd{X}_{\pm}\) and \(\dd{P}_{\pm}\) and project onto eigenstates corresponding to measurement outcomes \(I_{\pm}\), \(I'_{\pm}\). With the definition \(s_{\pm}=s_1\pm s_2\) we find
\begin{multline}
  \label{eq:51}
  \dd{\ket{\cond{\tilde{\psi}}}} = -\ii H_{\mathrm{eff}}\ket{\cond{\psi}}\dt\\
  +\sqrt{\upsilon/2}[s_+\dd{\widetilde{X}_+}+\ii s_-\dd{\widetilde{P}_-}]\ket{\cond{\psi}}\\
  +\sqrt{(1-\upsilon)/2}[\ii s_+\dd{\widetilde{X}_-}+ s_-\dd{\widetilde{P}_+}]\ket{\cond{\psi}}
\end{multline}
where \(\ket{\tilde{\psi}}\) is unnormalized. As all electromagnetic field modes are assumed to be in vacuum we find that the measurement processes have unit variance, \(\dd{\widetilde{X}_{\pm}}(t)^2=\dd{\widetilde{P}_{\pm}}(t)^2=\dt\), and that they are mutually uncorrelated, \ie{}, \(\dd{\widetilde{X}_+}\dd{\widetilde{X}_-}=\dd{\widetilde{X}_+}\dd{\widetilde{P}_+}=0\), etc. This can be shown by expressing \(c_i\) in terms of \(a\), \(b\) and using \Ito{} rules, where we have to take into account vacuum noise entering through the open ports of the second pair of beam-splitters (not explicitly introduced here). The homodyne currents are given by
\begin{subequations}
  \label{eq:52}
  \begin{align}
    I_+\dt&=\sqrt{\upsilon/2}\,\cond{\mean{s_++s_+^{\dagger}}}+\dW_+,\\
    I_-\dt&=\ii\sqrt{\upsilon/2}\,\cond{\mean{s_--s_-^{\dagger}}}+\dW_-,\\
    I'_+\dt&=\ii\sqrt{(1-\upsilon)/2}\,\cond{\mean{s_+ - s_+^{\dagger}}}+\dd{V_+},\\
    I'_-\dt&=\sqrt{(1-\upsilon)/2}\,\cond{\mean{s_-+s_-^{\dagger}}}+\dd{V_-}.
  \end{align}
\end{subequations}
where the Wiener increments \(\dW_{\pm}\) and \(\dd{V}_{\pm}\) obey a multiplication table corresponding to the one of \(\dd{\widetilde{X}}_{\pm}\) and \(\dd{\widetilde{P}}_{\pm}\). Following the derivation from \sref{SEC:homodyne-generic} with four uncorrelated homodyne measurements with non-unit efficiency we can derive the corresponding SME
\begin{multline}
  \label{eq:53}
  \dd{\cond{\rho}} = \mathcal{L}\cond{\rho}\dt+\sqrt{\upsilon/2} \left\{ \mathcal{H}[s_+]\cond{\rho}\dW_+ +\mathcal{H}[\ii s_-]\cond{\rho}\dW_- \right\}\\
  +\sqrt{(1-\upsilon)/2} \left\{ \mathcal{H}[\ii s_+]\cond{\rho}\dd{V}_+ +\mathcal{H}[s_-]\cond{\rho}\dd{V}_- \right\},
\end{multline}
with \(\mathcal{L}\rho = -\ii[H^\mathrm{(1)}_\mathrm{sys} + H^\mathrm{(2)}_\mathrm{sys},\rho] + \mathcal{D}[s_1]\rho + \mathcal{D}[s_2]\rho\).

These results can alternatively be derived in a similar spirit but in a more formal way within the framework of quantum networks, as for example presented in \cite{gough_series_2009}.

\subsubsection{Feedback master equation}
\label{sec-3-3-2}
In this entanglement swapping scheme all four homodyne currents, \(I_{\pm}\) (Bell measurement) and \(I'_{\pm}\) (stabilizing measurements), are fed back to \emph{both} systems. (For convenience we will in the following label the photo-currents by \(I_i\), \(i=1,\dots,4\), according to the light modes \(C_i\) they correspond to.) We again describe this by the operations \(\mathcal{K}[F_i]\rho I_i=-\ii [F_iI_i,\rho]\) (\(i=1,\dots,4\)), where \(F_i=F_i^{\dagger}\) now act on the combined Hilbert space of \(S_1+S_2\). Using the procedure from before it is straightforward to show that the corresponding FME can be written as
\begin{multline}
  \label{eq:54}
  \cond{\dot{\rho}}=-\ii \left[ H,\rho \right]-\frac{\ii}{2}\sum_{i=1}^4 \left[ s_i^{\dagger}F_i+F_i s_i,\cond{\rho}\right]\\
  +\sum_{i=1}^4 \left\{ \mathcal{D}[s_i-\ii F_i]+\frac{1-\eta}{\eta}\mathcal{D}[F_i] \right\},
\end{multline}
with \((s_i)_{i=1}^4=(\sqrt{\upsilon}s_+,\ii \sqrt{\upsilon}s_-,\ii \sqrt{1-\upsilon}s_+,\sqrt{1-\upsilon}s_-)\). Here we assumed that all detectors have the same efficiency \(\eta\).

\subsection{Optomechanical implementation}
\label{sec-3-4}

\subsubsection{Time-continuous teleporation}
\label{sec-3-4-1}
\label{SEC:teleport-om-deriv}
Here we derive the stochastic and feedback master equations for the optomechanical teleportation setup outlined in \sref{SEC:optom-teleport}, following the lines of \sref{SEC:teleport-generic} with modifications accommodating the optomechanical implementation. The one-dimensional electromagnetic field \(A\) couples to the cavity via the linear interaction \(H_{\mathrm{int}}=\ii \sqrt{\kappa}[\cc a^{\dagger}(t)-\cc^{\dagger}a(t)]\).
As before, \(A\) is assumed to be in the vaccuum state, while \(B\) is in a pure squeezed state.
In this section we refer to several different rotating frames: the frame of the driving laser rotating at \(\omega_0\) (which is our standard frame of reference), the squeezing frame which defines the central frequency for the squeezed input light at \(\omega_s\), and the local oscillator frame at \(\omega_{\mathrm{lo}}\) in reference to which all measurements will be made. We therefore have the relations
\begin{subequations}
  \label{eq:56}
  \begin{align}
    a(t)&=a_{\mathrm{lo}}(t)\ee^{-\ii\Delta_{\mathrm{lo}}t},\\
    b(t)&=b_{s}(t)\ee^{-\ii\Delta_{s}t}=b_{\mathrm{lo}}(t)\ee^{-\ii \Delta_{\mathrm{lo}}t},
  \end{align}
\end{subequations}
with the definitions \(\Delta_{\mathrm{lo}}=\omega_{\mathrm{lo}}-\omega_{0}\), \(\Delta_{s}=\omega_{s}-\omega_{0}\). The squeezed input state is then, in the squeezing frame, defined by the eigenvalue equation \(\left[ b_{s}(t)-\alpha b_{s}^{\dagger}(t) \right]\ket{M}_B = 0\) with \(\alpha=(N+M)/(N+M^{*}+1)\).
The Schr\"{o}dinger equation of the full system in the LO frame can be written as (neglecting for the moment the coupling to the mechanical bath as this can easily added in the end)
\begin{multline}
  \dd{\ket{\phi}}
  = -\ii H_{\mathrm{eff}}\ket{\phi}\dt +\sqrt{\kappa}\left(\dd{A}_{\mathrm{lo}}^{\dagger} +\alpha\dd{A}_{\mathrm{lo}}\right)\ee^{\ii \Delta_{\mathrm{lo}} t}\cc\ket{\phi}\\
  + \sqrt{\kappa}\left(\dd{B}_{\mathrm{lo}}\ee^{-\ii \delta t} -\alpha\dd{B}^{\dagger}_{\mathrm{lo}}\ee^{\ii\delta t}\right)\ee^{\ii \Delta_{\mathrm{lo}} t}\cc\ket{\phi},
\end{multline}
where \(\ket{\phi}\) is the state describing the complete system with an initial condition \(\ket{\phi_0} = \ket{\psi_0}_S\ket{\mathrm{vac}}_A\ket{M}_B\) and \(\delta=\Delta_{\mathrm{lo}}-\Delta_{\mathrm{s}}\). If we now choose \(\Delta_{s}=\Delta_{\mathrm{lo}}\), \ie{}, \(\delta=0\), we can rewrite this
as
\begin{align*}
  \dd{\ket{\phi}} &=\left[ -\ii H_{\mathrm{eff}}\dt+\sqrt{\kappa/2}\left( \mu\dd{X_+}+\ii\nu\dd{P_-} \right)\cc\,\ee^{\ii \Delta_{\mathrm{lo}}t} \right]\ket{\phi},
\end{align*}
where \(\dd{X_+}=\sqrt{1/2}(a_{\mathrm{lo}}+a_{\mathrm{lo}}^{\dagger}+b_{\mathrm{lo}}+b_{\mathrm{lo}}^{\dagger})\dt\) and \(\dd{P}_-=\ii\sqrt{1/2}(a_{\mathrm{lo}}-a_{\mathrm{lo}}^{\dagger}-b_{\mathrm{lo}}+b_{\mathrm{lo}}^{\dagger})\dt\), and \(\mu=1-\alpha\), \(\nu=1+\alpha\) as before. By comparing this to Schr\"{o}dinger equation \eqref{eq:37} we can deduce that the heterodyne Bell measurement at \(\omega_{\mathrm{lo}}\) is described by SME \eqref{eq:43} together with the expression for the measurement currents \eqref{eq:39} if we set \(s=\cc \ee^{\ii \Delta_{\mathrm{lo}}t}\). Thus the master equation
\begin{equation}
  \label{eq:57}
  \dd{\cond{\rho}}=\mathcal{L}\cond{\rho}\dt+\sqrt{\kappa}\mathcal{H}[(\mu\dW_++\ii\nu\dW_-)\cc \ee^{\ii \Delta_{\mathrm{lo}} t}]\cond{\rho}
\end{equation}
together with the output equations
\begin{subequations}
  \begin{align}
    \label{eq:58}
    I_+\dt&=\sqrt{1/2}\cond{\mean{\cc \ee^{\ii \Delta_{\mathrm{lo}} t}+\Hc{}}}\dt+\dW_+,\\
    \label{eq:59}
    I_-\dt&=\ii\sqrt{1/2}\cond{\mean{\cc \ee^{\ii \Delta_{\mathrm{lo}} t}-\Hc{}}}\dt+\dW_-
  \end{align}
\end{subequations}
provides us with a description of the conditional state of the full optomechanical system (including the cavity mode) conditioned on the heterodyne currents \(I_{\pm}\). What we eventually seek to obtain, however, is an effective description of the mechanical system only. In the weak coupling regime this can be achieved by adiabatically eliminating the cavity mode, which corresponds to a perturbative expansion in the small parameter \(g/\kappa\ll 1\). At the same time it will be important to keep $\kappa/\om$ and $\Dc/\om$ constant in order to capture the dynamical back-action effects of the cavity, which are crucial for a correct description of these systems. As this procedure is well covered in the literature \cite{doherty_feedback_1999}, we will only outline it briefly and point out the most important differences from earlier work. To be able to make the desired expansion we must first transform \eqref{eq:57} into the interaction picture defined by the free Hamiltonian \(H_0=\om \cm^{\dagger}\cm-\Dc \cc^{\dagger}\cc\),
\begin{multline}
  \label{eq:60}
  \dd{\cond{\tilde{\rho}}}=-\ii g\left[(\cc\ee^{\ii \Dc t}+\Hc{})(\cm\ee^{-\ii \om t}+\Hc{}),\cond{\tilde{\rho}}\right]\dt\\
  +\sqrt{\kappa}\mathcal{H}[(\mu\dW_++\ii\nu\dW_-)\cc \ee^{\ii (\Delta_{\mathrm{lo}}+\Dc) t}]\cond{\tilde{\rho}}\\
  +\kappa\mathcal{D}[\cc]\cond{\tilde{\rho}}\dt.
\end{multline}
(All operators marked with a tilde, \eg{}, \(\tilde{\rho}\), are defined with respect to this rotating frame.)
Following \cite{doherty_feedback_1999} one can show that the SME for the mechanical system (in the rotating frame at \(\om\)) can be written as
\begin{multline}
  \label{eq:61}
  \dd{\cond{\tilde{\rho}}^{(\m)}}=-\sqrt{2} g^2 \left[\tilde{x}_{\m}, \tilde{y}\cond{\tilde{\rho}}^{(\m)}-\cond{\tilde{\rho}}^{(\m)} \tilde{y}^{\dagger} \right]\dt\\
  + \sqrt{g^2\kappa} \mathcal{H}[(-\ii \mu \dW_++\nu\dW_-) \tilde{y}\ee^{\ii\Delta_{\mathrm{lo}}t}]\cond{\tilde{\rho}}^{(\m)},
\end{multline}
where \(\cond{\tilde{\rho}}^{(\m)}\) denotes the conditional state of the mechanical subsystem. Here we also defined \(\tilde{y}=\eta_-\cm\ee^{-\ii \om t}+\eta_+\cm^{\dagger}\ee^{\ii \om t}\) and \(\eta_{\pm} = [\kappa/2+\ii(-\Dc\pm\om)]^{-1}\).

Equation \eqref{eq:61} does not give rise to a valid Lindblad equation when averaged over all possible measurement trajectories as \(\tilde{y}\) is not a Hermitian operator. In order to get a consistent equation we apply a rotating wave approximation (RWA) to both the dynamics generated by the first commutator term and the measurement term. Let us take a closer look at the first term in \eqref{eq:60}: Plugging in the definitions of \(\tilde{x}_{\m}\) and \(\tilde{y}\) we find resonant terms of the form \(\cm \tilde{\rho}^{(\m)} \cm^{\dagger}\), \(\cm^{\dagger}\cm \tilde{\rho}^{(\m)}\), etc., and off-resonant terms oscillating at \(\ee^{\pm 2\ii \om t}\). The resonant terms have two effects: First they give rise to cooling and heating (see below), and second to a frequency shift of the mechanical resonance frequency (optical spring effect), yielding \(\om^{\mathrm{eff}}=\om+g^2\Im(\eta_++\eta_-)\). We have to account for this frequency shift by changing to a different rotating frame at \(\om^{\mathrm{eff}}\), which we still denote by \(\tilde{\rho}^{(\m)}\) for simplicity.

We can now introduce a time coarse graining in the form of \(\delta\cond{\tilde{\rho}}^{(\m)}=\int_t^{t+\delta t}\dd{\cond{\tilde{\rho}}^{(\m)}}\) which we apply to the resulting equation. We assume that it can be arranged such that \(\cond{\tilde{\rho}}^{(\m)}\) varies slowly on the timescale \(\delta t\) (and can thus be pulled out from under all time integrals), while we still average over many mechanical periods, \ie{}, \(\delta t\, \om^{\mathrm{eff}} \gg 1\). In the adiabatic regime the relevant system timescales are given by \(g^2/\kappa\) and \(\bar{n}\gamma\), the effective interaction strength and mechanical decoherence rate respectively. Hence we find that \(\delta t\) must fulfill \(\om^{\mathrm{eff}} \gg 1/\delta t \gg g^2/\kappa,\bar{n}\gamma\). Although equation \eqref{eq:61} is valid for any \(\Delta_{\mathrm{lo}}\) and \(\Dc\), the form of the resulting equation in RWA depends on the choice of \(\Delta_{\mathrm{lo}}\). As we illustrate in the main text we drive the optomechanical cavity on the blue sideband (\(\omega_0=\omega_{c}+\om\)), but want the LO to be resonant with the scattered photons (\(\omega_{\mathrm{lo}}=\omega_0-\om^{\mathrm{eff}}\)), and thus set \(\Delta_{\mathrm{lo}} = -\om^{\mathrm{eff}}\). For the first term in \eqref{eq:61} the RWA then simply amounts to dropping all terms oscillating with \(\ee^{\pm 2 \ii \om^{\mathrm{eff}} t}\) (as they are averaged out by the time coarse graining), which introduces an error of order \(1/(\delta t\,\om^{\mathrm{eff}})\). To treat the heterodyne measurement we introduce the coarse-grained noise increments \(\delta W_{\pm}^{(0)}=\int_t^{t+\delta t}\dW_{\pm}\), which obey \eqref{eq:16} if one replaces \(\dt\) with \(\delta t\). As we assumed that \(\delta t\) is small on the relevant timescales of the system in the interaction picture, we can now take the limit \(\delta t\rightarrow \dt\) (and thus also \(\delta \rho\rightarrow \dd{\cond{\rho}^{(\m)}}\), \(\delta W_{\pm}^{(0)}\rightarrow \dW_{\pm}^{(0)}\)). We find an effective SME for the mechanical system (valid for \(\Delta_{\mathrm{lo}}=-\om^{\mathrm{eff}}\) only),
\begin{multline}
  \label{eq:62}
  \dd\cond{\tilde{\rho}^{(\m)}}=
  \gamma_-\mathcal{D}[\cm]\cond{\tilde{\rho}^{(\m)}}\dt +
  \gamma_{+}\mathcal{D}[\cm^{\dagger}]\cond{\tilde{\rho}^{(\m)}}\dt\\
  +\sqrt{g^2\kappa/2}\;\mathcal{H}[(-\ii\mu \dW_{+}^{(0)} + \nu \dW_{-}^{(0)})\eta_+ \cm^{\dagger}]\cond{\tilde{\rho}^{(\m)}},
\end{multline}
where we added mechanical decoherence terms, and defined \(\gamma_- =\gamma(\bar{n}+1)+2g^2\mathrm{Re}(\eta_-)\) and \(\gamma_+ =\gamma\bar{n}+2g^2\mathrm{Re}(\eta_+)\). This equation generalizes the standard optomechanical MEQ from \cite{wilson-rae_theory_2007}. In principle there exist additional sideband modes centered at \(\pm 2 \om^{\mathrm{eff}}\), which in RWA are not correlated with \(W_{\pm}^{(0)}\), nor do are they entangled with the mechanical motion. We thus neglect them.

To apply feedback we have to extract the modes corresponding to the filtered noise processes \(W_{\pm}^{(0)}\) from the heterodyne currents \(I_{\pm}\). This can be achieved by applying the coarse-graining procedure from above to \eqref{eq:58} and \eqref{eq:59}, \ie{}, \(I_{\pm}^{(0)}=\int_t^{t+\delta t}I_{\pm}\dt\). Together with \(\cond{\mean{\cc}}=-\ii g\cond{\mean{y}}\), which results from the adiabatic elimination, we find
\begin{subequations}
  \label{eq:63}
  \begin{align}
    I_+^{(0)}\dt&\approx -\ii \sqrt{g^2\kappa/2}\mean{\eta_+\cm^{\dagger}-\Hc{}}\dt+\dW_+^{(0)},\\
    I_-^{(0)}\dt&\approx \sqrt{g^2\kappa/2}\mean{\eta_+\cm^{\dagger}+\Hc{}}\dt+\dW_-^{(0)},
  \end{align}
\end{subequations}
where we neglected contributions from higher sidebands, introducing corrections on the order \(1/(\delta t\, \om^{\mathrm{eff}})\).
With the identification \(s=-\ii\sqrt{g^2\kappa}\,\eta_+ \cm^{\dagger}\) the set of equations \eqref{eq:62}, \eqref{eq:63} is equivalent to the generic case discussed before. However, equation \eqref{eq:62} additionally contains decoherence terms due to the coupling to the mechanical environment (\(\gamma\bar{n}\mathcal{D}[\cm^{\dagger}]+\gamma(\bar{n}+1)\mathcal{D}[\cm]\)) and due to optomechanical back-action (\(2g^2\Re(\eta_-)\mathcal{D}[\cm]\)). For the choice \(F_+=-\sqrt{g^2\kappa}\eta_+(\cm+\cm^{\dagger})\) and \(F_-=\ii\sqrt{g^2\kappa}\eta_+(\cm-\cm^{\dagger})\) (where the prefactors are chosen to match the operator \(s\)), and after adding the appropriate decoherence terms, the FME for optomechanical teleportation can be written as
\begin{multline}
  \dot{\tilde{\rho}}^{(\m)}=\left\{ \gamma(\bar{n}+1)\mathcal{D}[\cm] + \gamma
    \bar{n}\mathcal{D}[\cm^{\dagger}] \right\} \tilde{\rho}^{(\m)}\\
  +\frac{4g^2}{\kappa}\left\{ (1+\epsilon) \mathcal{D}[\cm] + \frac{w_3}{\eta} \mathcal{D}[x_{\m}+p_{\m}]\right.\\
  + \left. \left(\frac{w_1-w_3}{\eta}-1\right)\mathcal{D}[p_{\m}]+\left(\frac{w_2-w_3}{\eta} -1\right)\mathcal{D}[x_{\m}]
  \right\}\tilde{\rho}^{(\m)},
\end{multline}
where \(\epsilon=[1+(4\om/\kappa)^2]^{-1}\) and we finally set \(\Dc=\om\). By applying the diagonalization procedure from \aref{APP:diag} we can bring this into the form \eqref{eq:17}.

\subsubsection{Time-continuous entanglement swapping}
\label{sec-3-4-2}
\label{SEC:om-swap-deriv}
In this section we derive the SME \eqref{eq:19} and FME \eqref{eq:21} which specify the generic case in \sref{SEC:swap-generic} for the optomechanical implementation. Again, the goal is to derive equations for the mechanical systems, which we obtain by adiabatic elimination of the cavity and subsequent application of a RWA. As before the Bell detection operates at the cavity frequency \(\omega_c\) detuned by \(\Delta_{\mathrm{lo}}=\omega_{\mathrm{lo}}-\omega_0\) with respect to the driving laser, and relations \eqref{eq:56} still apply. Following the logic from \aref{SEC:teleport-om-deriv} we thus define \(s_i=\sqrt{\kappa}\,\ee^{\ii \Delta_{\mathrm{lo}}t}c_{\lm,i}\), which we use together with the generic entanglement SME \eqref{eq:53} and FME \eqref{eq:52} as the starting point for our approximations. Going to the rotating frame with \(H_0=\sum_i (\om c_{\m,i}^{\dagger}c_{\m,i}-\Dc c_{\lm,i}^{\dagger}c_{\lm,i})\) and applying the adiabatic approximation procedure to \eqref{eq:53} leaves us with
\begin{multline}
  \label{eq:64}
  \dd{\cond{\tilde{\rho}}^{(\m)}}=-\sqrt{2} g^2 \sum_{i=1,2}\left[\tilde{x}_{\m,i}, \tilde{y}_i\cond{\tilde{\rho}}^{(\m)}-\cond{\tilde{\rho}}^{(\m)} \tilde{y}_i^{\dagger} \right]\dt\\
  +\sqrt{\frac{g^2\kappa\upsilon}{2}}\,  \mathcal{H}[(\tilde{y}_+\dW_++\ii \tilde{y}_-\dW_-)\ee^{\ii\Delta_{\mathrm{lo}}t}]\cond{\tilde{\rho}^{(\m)}}\\
  +\sqrt{\frac{g^2\kappa(1-\upsilon)}{2}}\,  \mathcal{H}[(\ii\tilde{y}_+\dd{V_+}+\tilde{y}_-\dd{V_-})\ee^{\ii\Delta_{\mathrm{lo}}t}]\cond{\tilde{\rho}^{(\m)}},
\end{multline}
where \(\tilde{y}_{\pm}=\tilde{y}_1\pm \tilde{y}_2\) with \(\tilde{y}_i=\eta_-c_{\m,i}\ee^{-\ii \om t}+\eta_+c_{\m,i}^{\dagger}\ee^{\ii \om t}\). To apply a time coarse graining \(\delta \cond{\rho}^{(\m)}=\int_t^{t+\delta t}\dd{\cond{\rho}^{(\m)}}\) we first change to the frame rotating with \(\om^{\mathrm{eff}}\) (taking into account the optical spring effect). If we choose \(\Delta_{\mathrm{lo}}=-\om^{\mathrm{eff}}\) we can drop the fast rotating terms in the first line of \eqref{eq:64}. For the measurement terms (second and third line) we again introduce \(\delta W_{\pm}^{(0)}\) and neglect any sideband modes. After taking the limit \(\delta t \rightarrow \dt\) we end up with
\begin{multline}
  \label{eq:65}
  \dd{\cond{\tilde{\rho}}^{(\m)}}=
  \gamma_- \{ \mathcal{D}[c_{\m,1}]+\mathcal{D}[c_{\m,2}]\} \cond{\tilde{\rho}^{(\m)}}\dt\\
  +\gamma_+ \{ \mathcal{D}[c_{\m,1}^{\dagger}]+\mathcal{D}[c_{\m,2}^{\dagger}]\}\cond{\tilde{\rho}^{(\m)}}\dt\\
  +\sqrt{g^2\kappa\upsilon/2}\,\eta_+ \mathcal{H}[c_{\m,+}^{\dagger}\dW_+^{(0)} + \ii c_{\m,-}^{\dagger} \dW_-^{(0)}]\cond{\tilde{\rho}^{(\m)}}\\
  +\sqrt{g^2\kappa(1-\upsilon)/2}\,\eta_+ \mathcal{H}[\ii c_{\m,+}^{\dagger}\dd{V_+^{(0)}} + c_{\m,-}^{\dagger} \dd{V_-^{(0)}}]\cond{\tilde{\rho}^{(\m)}}
\end{multline}
where we introduced \(c_{\m,\pm}=c_{\m,1}\pm c_{\m,2}\) and we added mechanical decoherence terms. We apply the same coarse graining to the measurement currents \eqref{eq:52} and find, by using \(s_i=\sqrt{\kappa}\ee^{\ii \Delta_{\mathrm{lo}}t}c_{\lm,i}\) and \(\cond{\mean{c_{\lm,i}}}=-\ii g\cond{\mean{y_i}}\),
\begin{subequations}
  \label{eq:66}
  \begin{align}
    I^{(0)}_+\dt&=-\ii\sqrt{g^2\kappa\upsilon/2}\,\cond{\mean{\eta_+c_{\m,+}^{\dagger}-\Hc{}}}+\dW_+^{(0)},\\
    I^{(0)}_-\dt&=\sqrt{g^2\kappa\upsilon/2}\,\cond{\mean{\eta_+c_{\m,-}^{\dagger}+\Hc{}}}+\dW_-^{(0)},\\
    I'^{(0)}_+\dt&=\sqrt{g^2\kappa(1-\upsilon)/2}\,\cond{\mean{\eta_+c_{\m,+}^{\dagger}+\Hc{}}}+\dd{V_+^{(0)}},\\
    I'^{(0)}_-\dt&=-\ii\sqrt{g^2\kappa(1-\upsilon)/2}\,\cond{\mean{\eta_+c_{\m,-}^{\dagger}-\Hc{}}}+\dd{V_-^{(0)}}.
  \end{align}
\end{subequations}
One can clearly see that equations \eqref{eq:65} and \eqref{eq:66} are equivalent to SME \eqref{eq:53} and measurement currents \eqref{eq:52} if we set \(s_{\pm}=\sqrt{g^2\kappa}\eta_+c_{\m,\pm}^{\dagger}\) and add appropriate decoherence terms. We can therefore use FME \eqref{eq:54} directly, and together with the choice \((F_1,F_2)=\sqrt{\upsilon}\beta(\ii c_{\m,+} - \ii c_{\m,+}^{\dagger}, c_{\m,-}+c_{\m,-}^{\dagger})\), \((F_3,F_4)=\sqrt{1-\upsilon}(c_{\m,+} + c_{\m,+}^{\dagger},\ii c_{\m,-} - \ii c_{\m,-}^{\dagger})\) we find equation \eqref{eq:21}.

\section{Conclusion}
\label{sec-4}

In this article we discuss different measurement-based feedback schemes which utilize entanglement as a resource in order to control the quantum state of mechanical systems. We derive and discuss in detail the dynamics of the optomechanical system under measurement and feedback, specifically the situations of feedback cooling, mechanical squeezing and generation of two-mode mechanical entanglement. The protocols are shown to be feasible in current optomechanical systems which operate in the strong-cooperativity regime.

\begin{acknowledgments}
  S.\,G.\ H.\ acknowledges helpful discussions with J.\ Schm\"o{}le.
  We thank support provided by the European Commission (ITN cQOM, iQUOEMS, and SIQS), the European Research Council (ERC QOM), the Austrian Science Fund (FWF): project numbers [Y414] (START), [F40] (SFB FOQUS), the Vienna Science and Technology Fund (WWTF) under Project ICT12-049, and the Centre for Quantum Engineering and Space-Time Research (QUEST). S.\,G.\ H.\ is supported by the Austrian Science Fund (FWF): project number [W1210] (CoQuS).
\end{acknowledgments}

\appendix

\section{Quantum stochastic calculus}
\label{sec-5}
\label{APP:qsc}
The joint unitary evolution of a system \(S\) and a white-noise electromagnetic field \(A\) can be described by a quantum stochastic differential equation in \Ito{}-form \cite{hudson_quantum_1984}
\begin{equation}
  \label{eq:67}
  \dd{U(t)} = \left[ -\ii H_{\mathrm{eff}}\dt+s \dd{A}^{\dagger}(t)-s^{\dagger}\dd{A}(t) \right]U(t),
\end{equation}
with \(U(0)=\mathds{1}\), where \(s\) is a system operator, and the effective Hamiltonian is given by \(H_{\mathrm{eff}}= H_{\mathrm{sys}}-\ii\frac{1}{2}s^{\dagger}s\), where \(H_{\mathrm{sys}}\) is the Hamiltonian describing the evolution of \(S\). \(A(t)\) and \(A(t)^{\dagger}\) are the bosonic annihilation and creation processes acting on the Fock space of the electromagnetic field. The increments \(\dd{A}\), \(\dd{A}^{\dagger}\) are forward-pointing, \(\dd{A}(t){:=}A(t+\dt)-A(t)\), and are (formally) connected to the singular field operators \(a(t)\) \cite{gardiner_input_1985} introduced in \sref{SEC:homodyne-generic} by \(\dd{A}(t)=a(t)\dt\). The definition of the increments leads to the property that they commute with \(U\) for equal times, \ie{}, \([U(t),\dd{A}(t)]=[U(t),\dd{A}^{\dagger}(t)]=0\).
If we assume that initially the electromagnetic field is in the vacuum state, the increments obey the multiplication rules
\begin{center}
  \begin{tabular}{c|ccc}
    \(\times\) & \(\dd{A}\) & \(\dd{A}^\dagger\) & \(\dt\)\\
    \hline
    \(\dd{A}\) & 0 & \(\dt\) & 0\\
    \(\dd{A}^\dagger\) & 0 & 0 & 0\\
    \(\dt\) & 0 & 0 & 0\\
  \end{tabular}
\end{center}
More generally two quantum stochastic processes \(X(t)\), \(Y(t)\) obey the \Ito{} product rule
\begin{equation}
  \label{eq:68}
  \dd{[X(t)Y(t)]}=[\dd{X(t)}]Y(t)+X(t)\dd{Y(t)}+\dd{X(t)}\dd{Y(t)},
\end{equation}
again with \(\dd{X}(t){:=}X(t+\dt)-X(t)\), etc. The standard chain rule is modified in a similar way. For a differentiable function \(f\), we have
\begin{equation}
  \label{eq:69}
  \dd{f(X(t))}=f'(X(t))\dd{X}(t)+\frac{1}{2}f''(X(t))\dd{X}(t)^2,
\end{equation}
which in particular leads to \(f(X(t+\dt))=f(X(t))+f'(X(t))\dX(t)+\tfrac{1}{2}f''(X(t))\dd{X}(t)^2\).
To convert between the \Ito{} and the Stratonovich formulation we can use the following approach \cite{gardiner_quantum_2004}. Consider the Stratonovich stochastic differential equation
\begin{equation}
  \label{eq:70}
  \mathbf{(S)}\qquad\dd{X}(t) = \mathcal{A}X(t)\dt+\mathcal{B}X(t)\dW(t),
\end{equation}
with linear operators \(\mathcal{A}\) and \(\mathcal{B}\), a Wiener process \(W(t)\) with \(\dW(t)^{2}=\dt\), and a initial condition \(X(0)\). This equation has the formal solution
\begin{equation}
  \label{eq:71}
  X(t)=\mathcal{T}\exp \left\{ \int_0^t[\mathcal{A}\dd{s}+\mathcal{B}\dW(s)] \right\}X(0),
\end{equation}
where \(\mathcal{T}\) denotes the time ordered product. We can now calculate the \Ito{} increment \(\dd{X}(t)=X(t+\dt)-X(t)\) and find
\begin{equation}
  \label{eq:72}
  \begin{aligned}
    \dd{X}(t)&=\left\{ \exp [\mathcal{A}\dd{s}+\mathcal{B}\dW(s)]-1 \right\}X(t)\\
    &=\left\{[\mathcal{A}+\tfrac{1}{2}\mathcal{B}^2]\dt+\mathcal{B}\dW(s)\right\}X(t),
  \end{aligned}
\end{equation}
where we expanded the exponential to second order and used \Ito{} rules. All stochastic differential equations in this manuscript are assumed to be in \Ito{} form unless noted otherwise [and denoted by an \((\mathbf{S})\)].

\section{Quantum LQG control}
\label{sec-6}
\label{APP:LQG}
In this section we briefly review the most important equations of quantum LQG control, following closely the presentation in \cite{edwards_optimal_2005}. Consider a Gaussian \(n\)-dimensional open quantum system coupling to \(m\) vacuum field channels, \(m'\leq m\) of which are subject to homodyne detection. (In the remainder of this section, we will assume that all channels are measured and thus \(m'=m\).\footnote{The case \(m'<m\) can be used to describe inefficient photo-detection (see \sref{SEC:homodyne-generic}) or decoherence channels which cannot be observed at all, \eg{}, phonon losses of a mechanical oscillator.}) The joint evolution of system plus field is then given by \eqref{eq:67}. The system's state conditioned on the outcomes of the homodyne measurements is described by the stochastic master equation (or quantum filter)
\begin{equation}
  \label{eq:73}
  \dd\cond{\rho} = -\ii [H,\cond{\rho}]\dt + \sum_{i=1}^m \mathcal{D}[L_i]\cond{\rho}\dt + \sum_{i=1}^{m'}\mathcal{H}[L_i]\cond{\rho}\dW_i,
\end{equation}
where \(\dW_i\) are Wiener processes with \(\dW_i\dW_j=\delta_{ij} \dt\) and the Hamiltonian is at most quadratic in the system's quadratures, which we collect into a column vector \(\vc{X}=(X_1,\dots,X_{2n})^{\trans}\). The canonical commutation relations can then be written as \([X_i,X_j]=\ii \mat{J}_{ij}\), where \(\mat{J}\) is an skew-symmetric real matrix. We can parametrize \(\vc{L}=(L_1,\dots,L_m)^{\trans}\) and \(H\) as \(\vc{L}=\mat{\Lambda}\vc{X}\) and
\begin{equation}
  \label{eq:74}
  H=\tfrac{1}{2}\vc{X}^{T}\mat{R}\vc{X}+\bigr[ \vc{X}^{\trans}\tilde{\mat{R}}\vc{u}(t)+\Hc{} \bigl],
\end{equation}
where \(\mat{R}\in \mathbb{R}^{n\times n}\) is symmetric, \(\tilde{\mat{R}} \in \mathbb{C}^{n\times m}\) and \(\vc{u}(t)\) is a \(m\)-dimensional input signal, which will later be used as a control input. We can describe the system in terms of a vector quantum Langevin equation and an output equation \cite{edwards_optimal_2005}
\begin{subequations}
  \label{eq:75}
  \begin{align}
    \label{eq:76}
    \dd{\vc{X}(t)}&=\bigr[ \mat{F}\vc{X}(t)+\mat{G}\vc{u}(t) \bigl]\dt+\dd{\vc{V}(t)},\\
    \dd{\vc{Y}(t)}&=\mat{H}\vc{X}(t)\dt+\bigr[\dd{\vc{A}(t)}+\dd{\vc{A}(t)}^{\dagger}\bigl],
  \end{align}
\end{subequations}
with the definitions \(\mat{F}=\mat{J}[\mat{R}+\Im(\mat{\Lambda}^{\dagger}\mat{\Lambda})]\), \(\mat{H}=\mat{\Lambda}+\mat{\Lambda}^{\dagger}\), \(\mat{G}=\mat{J}(\tilde{\mat{R}}+\tilde{\mat{R}}^{*})\), and \(\dd{\vc{V}}=\ii \mat{J}(\mat{\Lambda}^{\trans}\dd{\vc{A}}^{\dagger}-\mat{\Lambda}^{\dagger}\dd{\vc{A}})\), where \(\dd{\vc{A}}=(\dd{A_1},\dots,\dd{A_m})^{\trans}\). We assume the field is in the vacuum state \(\rho_{\mathrm{vac}}\), such that \(\dd{A_i}(t)\dd{A_j}(t)=\delta_{ij}\dt\). The measurement currents from the homodyne measurements are (formally) given by \(\vc{I}(t)=\dd{\vc{Y}}(t)/\dt\).

Using these definitions we can deduce the equations of motion for the conditional mean values \(\vc{\est{X}}=\tr{\vc{X}\cond{\rho}}\) and symmetric covariance matrix \(\mat{\est{\Sigma}}=\tr{\vc{X}\vc{X}^{\trans}\cond{\rho}}-\vc{\est{X}}\vc{\est{X}}^{\trans}\). We find \cite{edwards_optimal_2005}
\begin{subequations}
  \label{eq:77}
  \begin{align}
    \label{eq:78}
    \dd{\vc{\est{X}}}(t)&=\bigr[ \mat{F}\vc{\est{X}}(t)+\mat{G}\vc{u}(t) \bigl]\dt\notag\\
                        &\qquad+\mat{K}(t)\big[\dd{\vc{Y}}(t)-\mat{H}\vc{\est{X}}(t)\dt\big],\\
    \label{eq:79}
    \frac{\dd{}}{\dt}{\mat{\est{\Sigma}}(t)}&=\mat{F}\mat{\est{\Sigma}}(t)+\mat{\est{\Sigma}}(t) \mat{F}^{\trans}+\mat{N}\notag\\
                        &\qquad-\big[\mat{\est{\Sigma}}(t) \mat{H}^{\trans}+\mat{M}\big]\big[\mat{\est{\Sigma}}(t) \mat{H}^{\trans}+\mat{M}\big]^{\trans},
  \end{align}
\end{subequations}
where
\begin{subequations}
  \label{eq:80}
  \begin{align}
    \mat{K}(t)&=\mat{\est{\Sigma}}(t) \mat{H}^{\trans}+\mat{M},\\
    \mat{N}\dt&=\Re\bigl(\dd{\vc{V}}\dd{\vc{V}}^{\trans}\bigr)\notag\\
              &=\frac{1}{2}\mat{J}(\mat{\Lambda}^{\dagger}\mat{\Lambda}+\mat{\Lambda}^{\trans}\mat{\Lambda}^{*})\mat{J}^{\trans}\dt,\\
    \mat{M}\dt&=\Re\bigl(\dd{\vc{V}}(\dd{\vc{A}}+\dd{\vc{A}}^{\dagger})^{\trans}\bigr)\notag\\
              &=\frac{\ii}{2}\mat{J}(\mat{\Lambda}^{\trans}-\mat{\Lambda}^{\dagger})\dt.
  \end{align}
\end{subequations}
Equations \eqref{eq:77} together with \eqref{eq:80} are known as the \emph{Kalman--Bucy} filter in classical estimation theory \cite{kalman_new_1961}. Assuming a stable system \cite{wiseman_quantum_2009}, the steady-state solution of the conditional covariance matrix \(\vc{\est{\Sigma}}\) can be found by setting the right-hand side of \eqref{eq:79} to zero, and solving the resulting algebraic Riccati equation. If instead we are interested in the properties of the unconditional state, we can solve the Lyapunov equation obtained from \eqref{eq:79} by dropping the last term. [The resulting equation can also be obtained from \eqref{eq:76} by application of \Ito{} calculus.]

The goal of LQG control is to control a system in a way that minimizes a quadratic cost function. In this paper we only deal with the asymptotic control problem for \(t\rightarrow \infty\) as we are interested in the steady state of our systems. We therefore want to find a feedback strategy which minimizes the total cost \cite{bouten_separation_2008}
\begin{equation}
  \label{eq:81}
  \int_0^{\infty}\ev{h(\vc{X}(t),\vc{u}(t))}\dt,
\end{equation}
where we introduced \(\ev{\cdot}=\tr{\rho(0)\rho_{\mathrm{vac}}\,(\cdot)}\), the expectation value with respect to the initial state \(\rho(0)\) of the system and the vacuum state of the field. We choose \(h\) to be of the form
\begin{equation}
  \label{eq:82}
  h(\vc{X},\vc{u})=\vc{X}^{\trans}\mat{P}\vc{X}+\vc{u}^{\trans}\mat{Q}\vc{u},
\end{equation}
where \(\mat{P}\ge0\) and \(\mat{Q}>0\) are both real, symmetric matrices of appropriate dimensions. Under the assumption of certain stability conditions \cite{wiseman_quantum_2009} the optimal feedback signal is given by \cite{edwards_optimal_2005}
\begin{align}
  \label{eq:83}
  \vc{u}(t)&=-\mat{C}(t)\vc{\est{X}}(t),\\
  \mat{C}&=\mat{Q}^{-1}\mat{G}^{\trans}\sss{\mat{\Omega}},
\end{align}
with \(\sss{\mat{\Omega}}\) the solution of the algebraic Riccati equation
\begin{equation}
  \label{eq:84}
  \mat{F}^{\trans}\sss{\mat{\Omega}}+\sss{\mat{\Omega}}\mat{F}+\mat{P}-\sss{\mat{\Omega}}\mat{G}\mat{Q}^{-1}\mat{G}^{\trans}\sss{\mat{\Omega}} = 0.
\end{equation}
In \sref{SEC:cooling} we need to calculate the steady-state covariance matrix of a linear system including optimal feedback. This can be achieved by first noting that (due to the separation principle \cite{wiseman_quantum_2009}) we can write \eqref{eq:78} as
\begin{equation}
  \label{eq:85}
  \dd{\vc{\est{X}}}(t)=\bigr( \mat{F}-\mat{G}\mat{C} \bigl)\vc{\est{X}}(t)\dt+\mat{K}\dd{\widetilde{\vc{W}}},
\end{equation}
where \(\dd{\widetilde{\vc{W}}}\) is a Wiener process with \(\dd{\widetilde{\vc{W}}}(t)\dd{\widetilde{\vc{W}}}(t)^{\trans}=\mathds{1}_m\dt\) (the so-called innovations process). We also need that \cite{edwards_duality_2003,bouten_introduction_2007}
\begin{align}
  \label{eq:86}
  \Re\bigr(\ev{(\vc{X}(t)-\vc{\est{X}}(t))(\vc{X}(t)-\vc{\est{X}}(t))^{\trans}} \bigl)&=\mat{\est{\vc{\Sigma}}}(t),\\
  \label{eq:87}
  \ev{(\vc{X}(t)-\vc{\est{X}}(t))\vc{\est{X}}(t)^{\trans}}&=0,
\end{align}
where the first relation follows from the definition of \(\vc{\est{X}}\) and \(\vc{\est{\Sigma}}\), and the second from the orthogonality principle \cite{bouten_introduction_2007}. We therefore find
\begin{equation}
  \label{eq:88}
  \Re \bigr( \ev{\vc{X}(t)\vc{X}(t)^{\trans}} \bigl) = \mat{\est{\Sigma}}(t)+\ev{\vc{\est{X}}(t)\vc{\est{X}}(t)^{\trans}},
\end{equation}
where the equation of motion for the last term on the right-hand side \(\mat{\Xi}(t)=\ev{\vc{\est{X}}(t)\vc{\est{X}}(t)^{\trans}}\) can be deduced from \eqref{eq:85}, with a steady-state solution \(\sss{\mat\Xi}\) which fulfills
\begin{equation}
  \label{eq:89}
  (\mat{F}-\mat{GC})\sss{\mat{\Xi}}+\sss{\mat{\Xi}}(\mat{F}-\mat{GC})^{\trans}+\mat{K}\mat{K}^{\trans}=0.
\end{equation}
The steady-state solution of the symmetric covariance matrix of the controlled quantum system is thus given by
\begin{equation}
  \label{eq:90}
  \lim_{t\rightarrow\infty}\Re \bigr( \ev{\vc{X}(t)\vc{X}(t)^{\trans}} \bigl) = \sss{\mat{\est\Sigma}}+\sss{\mat{\Xi}}.
\end{equation}
Finally, we want to estimate the magnitude of the expected feedback signal. We quantify this by \(\ev{\vc{u}^{\trans}(t)\vc{u}(t)}\). In the steady state we find
\begin{equation}
  \label{eq:91}
  \ev{\vc{u}^{\trans}(t)\vc{u}(t)}=\ev{\vc{\est{X}}^{\trans}\mat{C}^{\trans}\mat{C}\vc{\est{X}}}=\tr{\mat{C}\sss{\mat{\Xi}}\mat{C}^{\trans}}.
\end{equation}

\section{Some basics on squeezed states}
\label{sec-7}
\label{APP:squeezed}
The field operator \(b(t)\) describing an ideal, white, squeezed light field fulfills
\begin{subequations}
  \begin{align}
    \label{eq:92}
    \mean{b^\dagger(t)b(t')}_{(N,M)}&=N\,\delta (t-t'),\\
    \mean{b(t)b(t')}_{(N,M)}&=M\,\delta (t-t'),
  \end{align}
\end{subequations}
with \(N>0\) and \(M\in \mathbb{C}\). For the quadratures \(x=(b+b^\dagger)/\sqrt{2}\) and \(p=-\ii(b-b^\dagger)/\sqrt{2}\) we therefore find
\begin{subequations}
  \label{eq:93}
  \begin{align}
    \mean{x^2}_{(N,M)}&=\tfrac{1}{2}(2N+1+M+M^*),\\
    \mean{p^2}_{(N,M)}&=\tfrac{1}{2}(2N+1-M-M^*),\\
    \tfrac{1}{2}\mean{xp+px}_{(N,M)}&= -\tfrac{\ii}{2}(M-M^*).
  \end{align}
\end{subequations}
For a physically meaningful state the uncertainty product must be \({(\Delta x)^2(\Delta p)^2-\tfrac{1}{4}\mean{xp+px}^2}\geq \tfrac{1}{4}\), which leads to \(|M|^2\leq N(N+1)\) (where equality is valid for a pure state).
In addition the pure squeezed state \(\ket{M}\) fulfills the eigenvalue equation
\begin{equation}
  \label{eq:94}
  \left[(N+M^*+1)b-(N+M)b^\dagger\right]\ket{M}=0.
\end{equation}
It follows that the displaced squeezed state \(\ket{M,\beta}=D(\beta)\ket{M}{:=}\exp( \beta b^\dagger - \beta^{*} b)\ket{M}\) fulfills the same equation if we make the replacement \(b\rightarrow b-\beta\).

\section{Diagonalization of non-Lindblad terms}
\label{sec-8}
\label{APP:diag}
In general the feedback master equations in \sref{SEC:meq} are not in Lindblad form as the prefactors of the operators \(\mathcal{D}\) can be negative. To cure this we can rewrite the non-unitary part of the evolution in terms of \(R=(x,p)^{\mathrm{T}}\) as \(\dot{\rho}=\sum_{ij}\Lambda_{ij}\left( R_i\rho R_j-\frac{1}{2} \rho R_j R_i-\frac{1}{2}R_j R_i\rho \right)\), where \(\Lambda\) is a Hermitian matrix. By virtue of the eigenvalue decomposition of \(\Lambda\) we can write \(\dot{\rho}=\sum_i\lambda_i \mathcal{D}[J_i]\rho\) with \(J_i=v_i \cdot R\), where \(\lambda_i\) and \(v_i\) \((i=1,2)\) are the eigenvalues and eigenvectors of \(\Lambda\) respectively.
\bibliography{TimeContinuousControl}

\end{document}